\documentclass[aps,prb,twocolumn]{revtex4} 

\usepackage{graphicx}
\usepackage{dcolumn}
\usepackage{bm}

\newcommand{\comment}[1]{}

\begin{document}

\title{More Reliable Measurements of the Slip Length
with the Atomic Force Microscope}


\author{Phil Attard}

\date{4--26 February, 2013. phil.attard1@gmail.com}

\begin{abstract}
Further improvements are made to
the non-linear data analysis algorithm
for the atomic force microscope
[P. Attard, arXiv:1212.3019v2 (2012)].
The algorithm is required
when there is curvature in the compliance region
due to photo-diode non-linearity.
Results are obtained for the hydrodynamic drainage force,
for three surfaces:
hydrophilic silica (symmetric, Si-Si),
hydrophobic dichlorodimethylsilane  (symmetric, DCDMS-DCDMS),
and hydrophobic octadecyltrichlorosilane (asymmetric, Si-OTS).
The drainage force was measured in
the viscous liquid di-n-octylphthalate.
The slip-lengths are found to be $3\,$nm for Si,
$2\,$nm for DCDMS,
and $2\,$nm for OTS,
with an uncertainty on the order of a nanometer.
These slip lengths are a factor of 4--15 times smaller
than those obtained from previous analysis of the same raw data
[L. Zhu et al., Langmuir, {\bf 27},  6712 (2011).
\emph{Ibid}, {\bf 28}, 7768 (2012)].
\end{abstract}


\maketitle

                \section{Introduction}

Measurements of the slip length are significant for three reasons.
First,
the hydrodynamic equations for any flow
can't be solved without specifying the boundary conditions,
and so whether a fluid sticks or slips during shear flow
at a solid surface is fundamental to the application
of hydrodynamics to the real world.
In turn, the extent that a fluid slips, if any,
effects quantitatively a range of physical phenomena
(e.g.\ flow rates in pores, drainage forces between particles,
lubrication of surfaces, pressure heads in microfluidic devices),
and so measuring, understanding, and controlling slip
could lead to new technologies, devices, and  industrial processes.

Second,
whereas equilibrium statistical mechanics is well-established
for elucidating static molecular structure (in bulk and at surfaces),
the same cannot be said for the non-equilibrium case,
either in general or for fluid flow.
Hence nanoscopic measurements of the slip length
yield fundamental information about
the structure of the fluid
and its interaction with the solid surface during shear flow.
Such data is valuable in its own right,
providing molecular-level insight into inhomogeneous flow
and perhaps enabling one to identify the specifically
non-equilibrium aspects of the way in which structure, flow,
and interaction are entwined at the molecular level.
In turn the measurements provide specific
motivation to develop non-equilibrium computer simulations,
and benchmarks against which those algorithms could be tested.

Third,
slip can be regarded as a correction to stick boundary conditions,
and as such it is a second order effect
that is a real challenge to measure experimentally
with any reliability or accuracy.
Obviously any small error in the measurement overall
translates into a large error in the slip length.
The literature abounds with examples where the reported slip lengths
vary by an order of magnitude for ostensibly the same system.
\cite{Neto05,Bocquet10}
(The present paper will present slip lengths that are a factor of 15 smaller
than those previously reported \emph{for the same data}.)
Hence credible measurements of the slip length
can only be obtained by improving the reliability and accuracy
of the measurement technique itself,
and this would have broad benefits
beyond the immediate application to slip.

On this last point,
the present paper uses the atomic force microscope
to measure the slip length via the analysis of the hydrodynamic drainage force.
The present results represent the third generation
of improvements to atomic force microscopy
that have been motivated
by the desire for reliable measurements of the slip length.
Most of the new algorithms and procedures are general in nature
and can be directly applied to other forces
measured with the atomic force microscope.
In the present case, where the sought for slip length is a second order effect,
the new procedures have proved essential for reliable results.
In other cases, where one might be measuring a first order effect,
the improvements are still worthwhile
because they reduce the quantitative error of the measurements
by about an order of magnitude
while still remaining simple enough
for the data analysis to be carried out by a spreadsheet.

\begin{figure}[t!]
\centerline{
\resizebox{8.5cm}{!}{ \includegraphics*{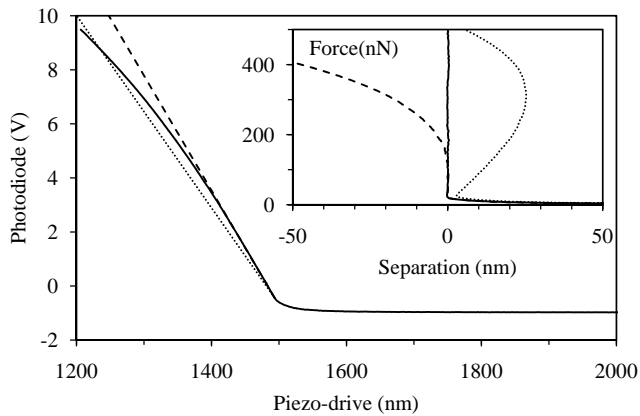} } }
\caption{\label{Fig:VvsZp}
The raw photo-diode voltage versus the piezo-drive displacement.
The solid curve is measured extension data,
the dashed line gives the tangent at first contact,
and the dotted line gives the average slope in contact.
The inset shows the analysed force versus separation,
with the solid curve resulting from the non-linear analysis
and the dashed and dotted curves resulting from the conventional linear analysis
using the first contact slope and the average slope, respectively.
The source of the measured data is Ref.~\onlinecite{Zhu12},
which may be consulted for the experimental details.
}
\end{figure}

In Appendix \ref{Sec:AppA},
the various improvements in force measurement and data analysis
with the atomic force microscope are summarized.
The focus is on those developments stimulated by the
challenge of accurate and reliable measurements of the slip length,
although of course many of these improvements
are more generally applicable.
As discussed in the appendix,
the most recent series of improvements concern
the non-linear analysis of the measured force data.\cite{Attard12}
The need for such an algorithm arises
when the compliance region has curvature,
as is shown in Fig.~\ref{Fig:VvsZp}.
The non-linearity is due to the response of the photo-diode,
most likely arising from non-uniformity in intensity and width
of the light beam moving across the split photo-diode.
In Ref.~\onlinecite{Attard12},
an algorithm was given based on a polynomial of best fit
to the contact region,
from which the angular and vertical deflection of the cantilever
in the non-contact region can be obtained.
The algorithm was applied to measured raw data
for the drainage force in the Si-DOPC-Si system.

This paper presents further improvements
to the non-linear data analysis algorithm
that are applicable to general atomic force microscopy.
The modified analysis is applied to two new systems,
DCDMS-DOPC-DCDMS and Si-DOPC-OTS,
as well as re-analysing the Si-DOPC-Si system.

One improvement concerns replacing an extrapolation
of the polynomial fit by an inherently more reliable interpolation.
This can be explained as follows.
The voltage measured on extension can be split into two ranges,
in contact $V \ge V_\mathrm{c,ext}$,
and prior to contact $V \le V_\mathrm{c,ext}$
(because the drainage force is repulsive),
where $V_\mathrm{c,ext}$ is the voltage at initial contact.
This means that in order to obtain the pre-contact drainage force
on extension from the measured voltage,
one has to use an extrapolation of the non-linear fit made in contact.
This extrapolation can lead to small but unacceptable errors.
In contrast,
on retraction,
in contact the voltage is $V \ge V_\mathrm{c,ret}$,
and after contact $V \agt V_\mathrm{c,ret}$
(because the drainage force is attractive),
the voltage at final contact, $V_\mathrm{c,ret}$,
is strictly less than the voltage
measured on the whole range of extension.
This means that one can use the non-linear fit
to the retraction data in contact
to obtain both the extension and the retraction out of contact force
by interpolation rather than extrapolation.
This gives a small but significant improvement in the drainage results.

A second improvement is that a least squares fit algorithm
has been developed to obtain the cantilever spring constant
and effective drag length
from the measured data at large separations.
By automating this procedure,
a source of possible human bias is eliminated
and the time required to analyze each force curve
is significantly reduced.

A third improvement was also explored,
which in some systems can be important,
but which makes negligible difference
for the atomic force microscope data analyzed here.
This is variable cantilever drag,
which arises from the fact that the hydrodynamic drag on the cantilever
decreases as the cantilever bends in response to the drainage force.
This effect was not taken into account in
the theoretical forces that were fitted
to the non-linearly analyzed data.\cite{Attard12}
In earlier linear analysis,
variable drag was found to be significant
for weak cantilevers ($k_\mathrm{eff} \alt .1\,$N/m),
but negligible for stiff ones($k_\mathrm{eff} \agt 1\,$N/m).
\cite{Zhu11a,Zhu11b,Zhu12a}
Since the spring constant obtained with the non-linear analysis
was relatively large,
$k_\mathrm{eff} = 1.68\,$N/m,\cite{Attard12}
it was assumed that it would be acceptable in the first instance
to neglect variable drag.
In the present work this assumption is checked
by performing variable drag calculations for the experimental conditions,
and it is found that it does indeed have negligible effect.
However in the process of performing the check,
a more robust numerical algorithm was developed
that gives more reliable results
and that allows the retract data to be calculated as well.
This algorithm has some intrinsic interest and is included
in Appendix B,
which sets out in detail the non-linear data analysis algorithm.

For an independent experimental test of the present protocols,
the procedure used in Ref.~\onlinecite{Zhu12} is used here as well.
In that case three sets of measurements were performed
for the drainage force in the liquid DOPC (di-n-octylphthalate):
Si-Si, DCDMS-DCDMS, and Si-OTS,
where Si is a silicon wafer with a native silicon oxide layer,
DCDMS is a dichlorodimethylsilane self-assembled monolayer
prepared on the same type of silicon wafer from the vapor phase,
and OTS is an octadecyltrichlorosilane self-assembled monolayer
also prepared on a silicon wafer.
In water, Si is hydrophilic with a contact angle close to zero,
whereas DCDMS is hydrophobic with an advancing contact angle
of 109$^\circ$,\cite{Zhu12}
and OTS is also hydrophobic with contact angle of 112$^\circ$.\cite{Zhu11b}
In DOPC
the contact angle is 21$^\circ$ for Si,\cite{Zhu12}
48$^\circ$ for DCDMS,\cite{Zhu12}
and 45$^\circ$ for OTS.\cite{Zhu11b}
Given the similarity of the two hydrophobic monolayers,
one would expect them to have similar slip lengths,
and the two slip lengths fitted for the symmetric systems
ought also fit the asymmetric system.
This represents a test of the reliability of the measurement protocol.




%
\section{Results}
%

\begin{table}[h!]
\caption{\label{tb:I}
Parameters and results for three series of
drainage force measurements.$^{*}$}
\begin{tabular}{l c c c }
\hline
                           &    Si-Si     & DCDMS-DCDMS  & Si-OTS \\
No.\ Meas.\                & 9            & 11           & 10        \\
Cantilever$^\dag$          & C            & A            & F        \\
$ L_0 $ ($\mu$m)           & 110          & 90           & 230      \\
$ R $   ($\mu$m)           & 10.11        & 9.28         & 10.14     \\
$ k_0$  (N/m)              & $1.38\pm .06$& $1.78\pm .10$&  $1.20\pm .07$\\
$k_\mathrm{eff}$ (N/m)     & $1.69\pm .08$& $2.22\pm .12$&  $1.34\pm .08$\\
$L_\mathrm{drag}$ ($\mu$m) & $85\pm 8$    & $71\pm 5$    & $162\pm 5$ \\
$A$ (J$\,$m$^{-2}$)        & $1 \times 10^{-21}$
& $5\times 10^{-20}$ & $1\times 10^{-20}$  \\
$b$  (nm)                  & 3            & 2            & 3,2 \\
\hline
\end{tabular}
\begin{flushleft}
$^*$All measurements were performed in di-n-octylphthalate (DOPC),
with viscosity $\eta =$ 50--54\,mPa\,s.
The average and standard deviation over different drive velocities are given.
Here $L_0$ is the length of the cantilever,
$R$ is the radius of the colloid probe,
the $k$ are cantilever spring constants (see text),
$L_\mathrm{drag}$ is the effective drag length,
$A$ is the Hamaker constant,
and $b$ is the slip length. \\
$^\dag$NSC12, tipless, rectangular (Mikromasch).\\

\end{flushleft}
\end{table}

Raw data for three systems were analysed:
Si-Si, DCDMS-DCDMS, and Si-OTS.
These are summarised in Table~\ref{tb:I}.
The tilt angle of the cantilever was $-11^\circ$.
The cantilever length quoted in the table is
$L_0 = L_0^* - 2R$,
where $L_0^* $ is the nominal length quoted by the manufacturer.
This recipe accounts approximately for the fact that the colloid probe
is mounted slightly back from the leading edge of the cantilever.

There is a one-to-one relationship between
the intrinsic cantilever spring constant $k_0$
and the effective force measuring spring constant
$k_\mathrm{eff}$,
which is given in Eq.~(\ref{Eq:keff}) below.
Usually both are quoted below
even though this is redundant.
In brief, the effective spring constant
gives the surface force from
the vertical deflection of the contact point,
$  F_z = k_\mathrm{eff} z_\mathrm{c}$,
whereas the intrinsic cantilever spring constant
gives the hypothetical force acting normal to the axis
of the cantilever that would cause
a deflection of the cantilever also normal to its axis,
$F_\perp = k_0 \zeta$.
Most calibration techniques yield $k_0$
and this is what is usually quoted and used in papers
even though it is $k_\mathrm{eff}$
that is required for the quantitative analysis
of atomic force microscopy.
In the present paper
the spring constant
was determined by a least squares fit of the drainage and drag force
to the measured data at large separations.
The error is about one third of the standard deviation shown
in Table~\ref{tb:I}.

In the figures below the vertical deflection of the contact point
$z_\mathrm{c}$ is generally given.
In previous papers this was called the vertical deflection of the tip
and was denoted $z_\mathrm{t}$.\cite{Zhu11a,Attard12}
It is more precise to call it the vertical deflection of the contact point,
and to instead write $\zeta_L C_0$
for the vertical component of the deflection of the end of the cantilever
(see \S\ref{Sec:Analysis}).
In contact, $\Delta z_\mathrm{c} = -\Delta z_\mathrm{p}$,
where $z_\mathrm{p}$ is the piezo-drive position.
As just mentioned,
the surface force is related to this by $F_z = k_\mathrm{eff} z_\mathrm{c}$.

In the deflection-separation curves given below,
the constant cantilever drag force has been subtracted.
The theoretical calculations also have constant drag subtracted.
Calculations have been made with constant and with variable
cantilever drag.
For the three systems treated below,
the effect of variable drag is negligible.
The variable drag algorithm used in places below
improves upon the original\cite{Zhu11a} in that it is more robust
(in fact, it is more reliable than the spread sheet algorithm
for constant drag),
it includes tilt and torque,
and results are obtained for adhesion and retraction.

\subsection{Si-Si}

\begin{figure}
\centerline{
\resizebox{8.5cm}{!}{ \includegraphics*{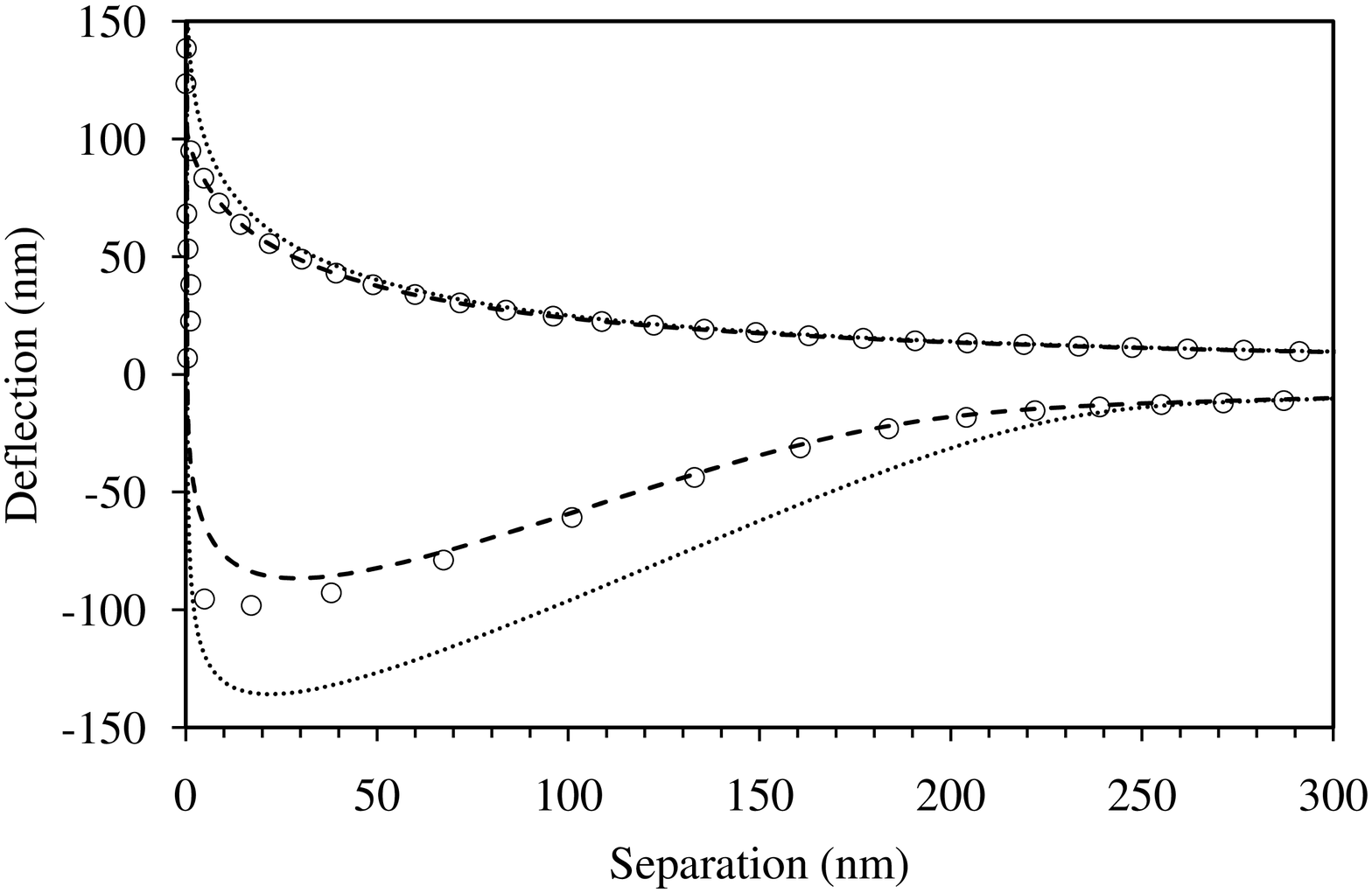} } }
\caption{\label{Fig:zvsh-SiSi50}
Vertical cantilever deflection $z_\mathrm{c}$ versus separation $h$
at a drive rate of $\dot z_\mathrm{p} = 50\,\mu$m$\,$s$^{-1}$
for Si-Si (see Table~\ref{tb:I}).
The effective cantilever spring constant is $k_\mathrm{eff} = 1.69\,$N/m.
The circles are the analysed experimental measurement
(every tenth point plotted),
the dotted curve is the calculation with stick
boundary conditions, $b=0$\,nm,
and the dashed curve is the slip calculation, $b=3$\,nm.
The source of the raw measured data is Ref.~\onlinecite{Zhu12}.
}
\end{figure}

Figure~\ref{Fig:zvsh-SiSi50}
shows atomic force microscope measurements of the drainage force for Si-Si.
The vertical cantilever deflection is shown,
from which the force may be obtained as $F_z = k_\mathrm{eff} z_\mathrm{c}$.
The raw data was analysed using the non-linear procedures described
in Appendix \ref{Sec:AppB}.
It can be seen that the drainage force is repulsive on extension (approach)
and attractive on retraction (retreat).
At large separations there is almost complete overlap between
the measured data and the two calculated curves.
At small separations, it can be seen that the stick theory
significantly overestimates the magnitude of the measured drainage force
on both extension and retraction.
The slip theory with  $b=3$\,nm shows an almost perfect fit
on extension all the way into contact.
The agreement is less good on retraction immediately pulling out of contact,
but by about 40$\,$nm  the measured and calculated deflection converge.

The same raw data was non-linearly analysed in Ref.~\onlinecite{Attard12}.
The difference in the two is that
the present analysis uses the non-linear fit for the voltage
in contact on retraction to obtain the cantilever deflection angle
out of contact on extension.
Also, here the cantilever spring constant was obtained
by a least squares fit of stick theory to the large separation data,
$h \in [0.9,3.7]\,\mu$m.
This was done at each drive velocity
and the result averaged over all  velocities.
In Ref.~\onlinecite{Attard12} the fit was performed by eye.
The value of the spring constant obtained previously
was $k_\mathrm{eff} = 1.68\,$N/m,\cite{Attard12}
whereas the present least squares fit gave $k_\mathrm{eff} = 1.69\pm .08\,$N/m,
which suggests that the author has a very good eye.
The slip length obtained here,  $b=3$\,nm, is the same as
that obtained previously.\cite{Attard12}
In the original linear analysis of the atomic force microscope data,
the spring constant fitted was $k_\mathrm{eff} = 1.5\,$N/m,
and the slip length fitted at low shear rates was $b_0=10$\,nm.\cite{Zhu12}

\begin{figure}
\centerline{
\resizebox{8.5cm}{!}{ \includegraphics*{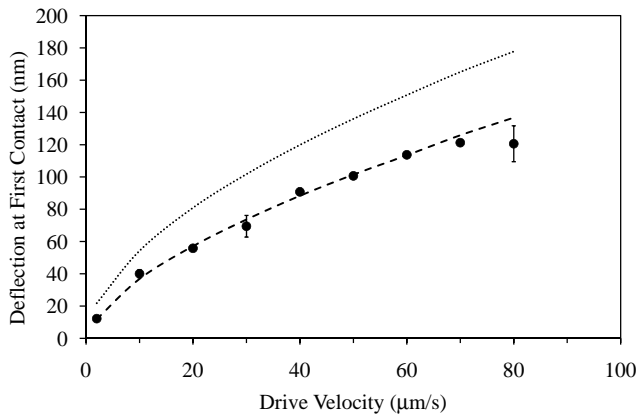} } }
\caption{\label{Fig:1stct-SiSi}
Deflection $z_\mathrm{c}$ at first contact
versus drive velocity for Si-Si.
The symbols are the analysed measured data
(raw data from Ref.~\onlinecite{Zhu12}),
the dotted curve is the stick calculation $b=0$\,nm,
and the dashed curve is the slip calculation with  $b=3$\,nm,
with other parameters as in Table~\ref{tb:I}.
The bars are a crude estimate of the error arising
from the uncertainty in identifying first contact.
}
\end{figure}

The slip theory fits the measured data on extension
almost perfectly in Fig.~\ref{Fig:zvsh-SiSi50},
and this level of agreement is reasonably typical of all
the forces analysed for Si-Si.
This can be seen in Fig.~\ref{Fig:1stct-SiSi},
where the cantilever deflection at first contact on extension is shown.
In the case of the experiments,
there can be an uncertainty
(on the order of several nanometers in the drive distance)
in deciding exactly where first contact occurs,
and it is crudely estimated that this gives
an error on the order of 2--10$\,$nm in the deflection.
Alternatively, the lack of smoothness in the measured data points
in Fig.~\ref{Fig:1stct-SiSi} also gives a guide to the error
in the deflection at first contact.
Within this error,
it can be seen that a slip length of $b=3$\,nm
gives a  deflection at first contact in quantitative agreement
with the measured one over the whole range of drive velocities.
The stick theory results shown in the figure
correspond to $b=0$\,nm and overestimate the deflection
by about 35\% at $\dot z = -50\,\mu$m/s.
From this one can conclude
that any variation in the slip length
by more than a fraction of a nanometer will lead to a significantly worse fit
than has been obtained for  $b=3$\,nm.

The van der Waals force has very little effect on the force on extension.
For example, at a drive velocity of  $\dot z = -50\,\mu$m/s,
the calculated  deflection at first contact for  $b=3\,$nm
and a Hamaker constant of $A= 1.3\times 10^{-21}\,$J$\,$m$^{-2}$
is 101.4$\,$nm.
Changing the Hamaker constant to $A= 5\times 10^{-21}\,$J$\,$m$^{-2}$,
the deflection becomes 100.9\,nm.

\begin{figure}
\centerline{
\resizebox{8.5cm}{!}{ \includegraphics*{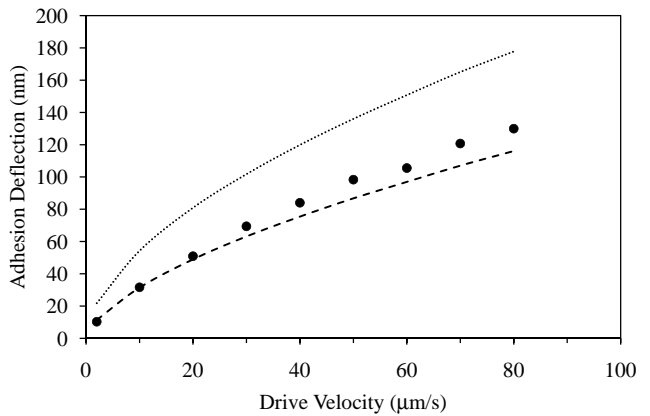} } }
\caption{\label{Fig:Ahsn-SiSi}
Adhesion deflection $z_\mathrm{c}$ versus drive velocity for Si-Si.
The symbols, curves and parameters are as in the preceding figure.
The adhesive tension is $F_z = k_\mathrm{eff} z_\mathrm{c}$.
The calculations use
a Hamaker constant of $A=1.3\times10^{-21}\,$J$\,$m$^{-2}$
and zero force position of $z_0 = 0.53\,$nm.
}
\end{figure}

Figure \ref{Fig:Ahsn-SiSi} shows the adhesion deflection,
which is defined as the negative of the minimum deflection that occurs
on retraction.
This is easier to obtain and shows less variability than
the deflection at last contact.
In general terms obtaining the adhesion reliably in atomic force microscopy
is extremely challenging.
This is because the jump out of contact is a catastrophic event
and it is exceedingly sensitive to external vibrations,
surface roughness, elastic deformation,
and to sliding and rolling of the probe,
and to peeling of the contact region.
The present drainage force appears to be an exception,
since there is quite good agreement between theory
and measurement in this case.
Although the instant of the jump out is variable,
the minimum in the force curve, which is the maximum tension,
is quite stable and appears to be largely determined by the drainage force.

It can be seen in  Fig.~\ref{Fig:Ahsn-SiSi}
that stick theory significantly overestimates the adhesion,
whereas slip theory with $b=3\,$nm is relatively accurate.
From the high velocity data one might conclude that
$b=3\,$nm is an upper bound on the slip length.

Although the actual pull-off deflection at last contact
can be sensitive  to the van der Waals force,
the adhesion deflection as defined here is less so.
For example, at a drive velocity of  $\dot z = -50\,\mu$m/s,
the measured adhesion deflection is 98.3$\,$nm.
The calculated adhesion deflection
for a Hamaker constant of $A= 1.3\times 10^{-21}\,$J$\,$m$^{-2}$
and a slip length of $b=3\,$nm is 86.7$\,$nm,
and that for stick, $b=0\,$nm is 136.0$\,$nm.
Changing the Hamaker constant to $A= 5\times 10^{-21}\,$J$\,$m$^{-2}$
for the slip length of $b=3\,$nm,  the deflection becomes 88.9$\,$nm.

\subsection{DCDMS-DCDMS}

\begin{figure}
\centerline{
\resizebox{8.5cm}{!}{ \includegraphics*{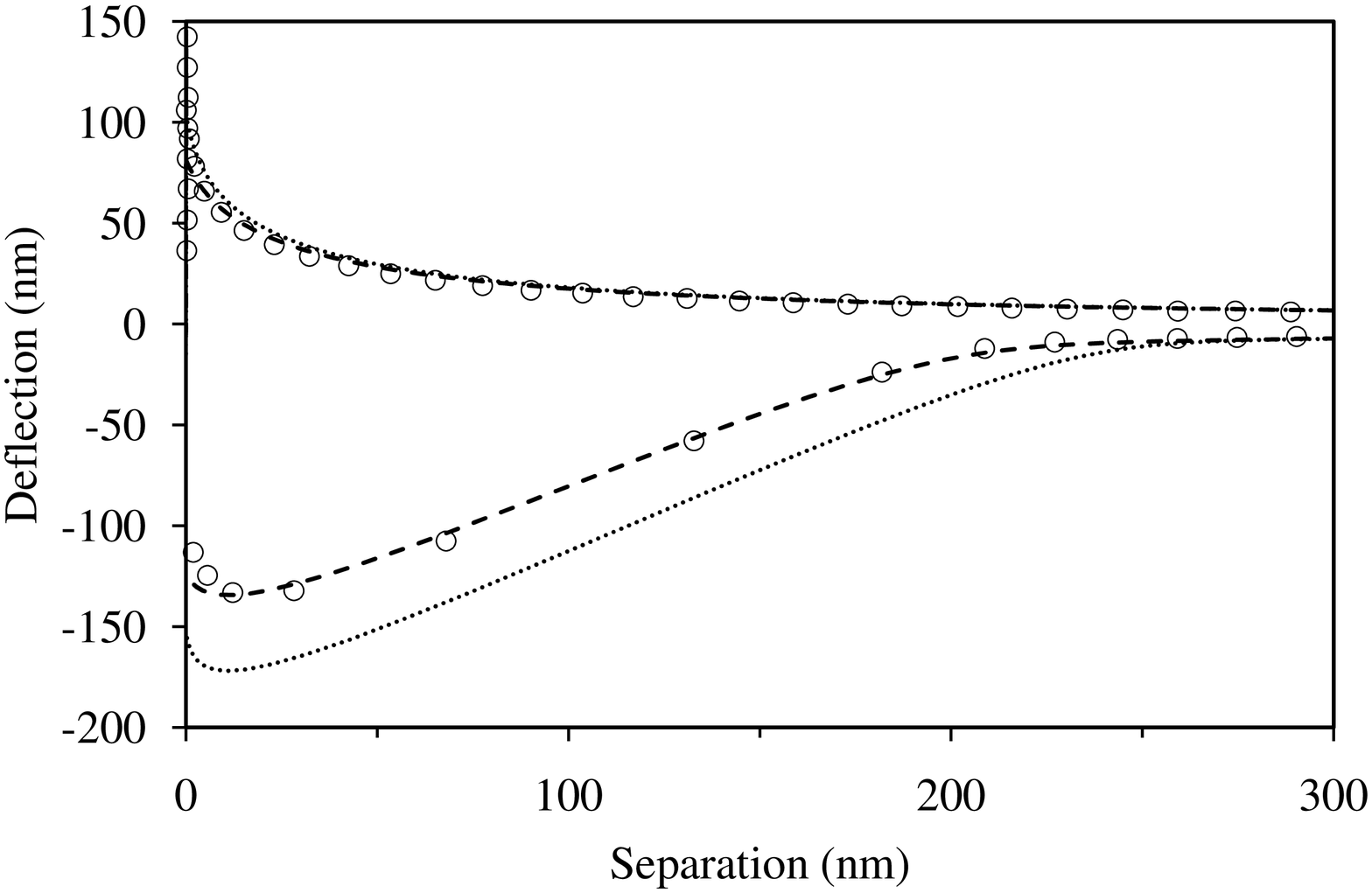} } }
\caption{\label{Fig:zvsh-DCDC50}
Vertical cantilever deflection $z_\mathrm{c}$ versus separation $h$
at a drive rate of $\dot z_\mathrm{p} = 50\,\mu$m$\,$s$^{-1}$
for DCDMS-DCDMS (see Table~\ref{tb:I}).
The effective cantilever spring constant is $k_\mathrm{eff} = 2.22\,$N/m.
The circles are the analysed experimental measurement
(every tenth point plotted),
the dotted curve is the stick calculation, $b=0$\,nm,
and the dashed curve is the slip calculation, $b=2$\,nm.
The source of the raw measured data is Ref.~\onlinecite{Zhu12}.
}
\end{figure}

Analysed atomic force microscope results
for the vertical deflection of the cantilever
as a function of separation for DCDMS-DCDMS
are shown in Fig.~\ref{Fig:zvsh-DCDC50}.
Again it can be seen that the stick theory overestimates the magnitude
of the force at small separations on both extension and retraction.
In contrast, the slip theory with $b=2$\,nm
fits the measured results on extension down to several nanometers
from first contact.
It is also in surprisingly good agreement
for the  measured retraction force.

\begin{figure}
\centerline{
\resizebox{8.5cm}{!}{ \includegraphics*{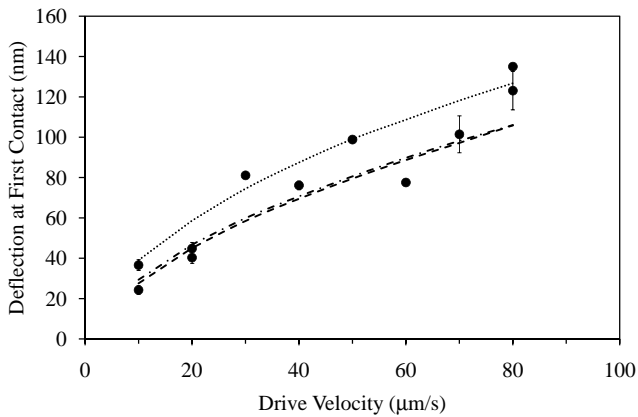} } }
\caption{\label{Fig:1stct-DCDC}
Deflection $z_\mathrm{c}$ at first contact
versus drive velocity for DCDMS-DCDMS.
The symbols are the analysed measured data
(raw data from Ref.~\onlinecite{Zhu12}),
the dotted curve is the stick calculation $b=0$\,nm
and a Hamaker constant of $A=5\times10^{-20}\,$J$\,$m$^{-2}$,
the dashed curve is the slip calculation with  $b=2$\,nm
and  $A=5\times10^{-20}\,$J$\,$m$^{-2}$,
and the dash-dotted curve is the slip calculation with  $b=2$\,nm
and $A=1\times10^{-20}\,$J$\,$m$^{-2}$.
The other parameters are as in Table~\ref{tb:I}.
The bars are a crude estimate of the error arising
from the uncertainty in identifying first contact.
}
\end{figure}

The extension deflection at first contact
is shown  in Fig.~\ref{Fig:1stct-DCDC}
as a function of drive velocity,
including a number of repeat measurements.
There appears to be a larger scatter in the measured data
in this case compared to Si-Si analysed in Fig.~\ref{Fig:1stct-SiSi}.
The crude error estimate appears to be on the low side.
Calculated results for $A=5\times10^{-20}\,$J$\,$m$^{-2}$,
and  for $A=1\times10^{-20}\,$J$\,$m$^{-2}$
are shown to give an idea of the sensitivity to the Hamaker constant.
Most of the measured data lie between $b=0$\,nm and $b=2$\,nm.
It is a little unexpected that the slip length
for the hydrophobic low energy surface DCDMS
should be less than the slip length for the hydrophilic
high energy surface Si.

\begin{figure}
\centerline{
\resizebox{8.5cm}{!}{ \includegraphics*{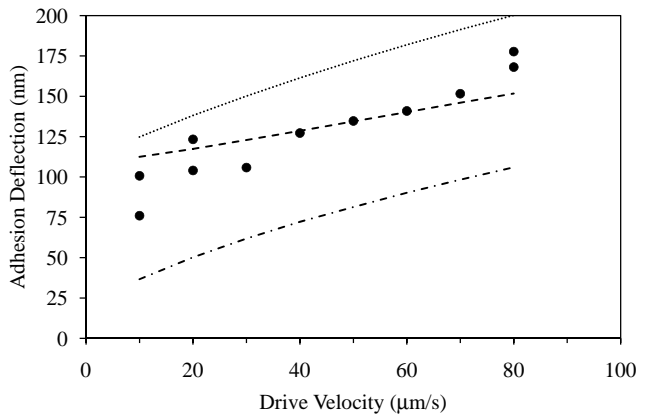} } }
\caption{\label{Fig:Adhsn-DCDC}
Adhesion deflection $z_\mathrm{c}$ versus drive velocity for DCDMS-DCDMS.
The symbols, curves, and parameters are as in the preceding figure.
The adhesive tension is $F_z = k_\mathrm{eff} z_\mathrm{c}$.
}
\end{figure}

Figure \ref{Fig:Adhsn-DCDC}
shows the adhesion deflection as a function of drive velocity,
again including repeat measurements.
Again most of the measured data lie between $b=0$\,nm and $b=2$\,nm.
The Hamaker constant that was used,  $A=5\times10^{-20}\,$J$\,$m$^{-2}$,
is forty times larger that was used for Si-Si.
It can be seen that a Hamaker constant of  $A=1\times10^{-20}\,$J$\,$m$^{-2}$
significantly underestimates the adhesion.
(One might argue from the low velocity data
that a value of $A=4\times10^{-20}\,$J$\,$m$^{-2}$
would give a better fit.
One might also argue that  a smaller slip length would better fit
the high velocity data.)
To be concrete,
at a drive velocity of $\dot z_\mathrm{p} = 50\,\mu$m$\,$s$^{-1}$,
the measured adhesion deflection is 134.7\,nm.
For $b=2$\,nm and  $A=5\times10^{-20}\,$J$\,$m$^{-2}$,
the calculated adhesion is 133.8\,nm,
and for  $b=2$\,nm and  $A=1\times10^{-20}\,$J$\,$m$^{-2}$,
it is 81.3\,nm.
These calculations were performed with the variable drag algorithm.
Using constant drag instead,
for $b=2$\,nm and  $A=5\times10^{-20}\,$J$\,$m$^{-2}$,
the calculated adhesion is 134.9\,nm,
which is a minor change.
(In this case the deflection at first contact
increases by 0.9\,nm using constant rather than variable drag.)

\subsection{Si-OTS}

\begin{figure}
\centerline{
\resizebox{8.5cm}{!}{ \includegraphics*{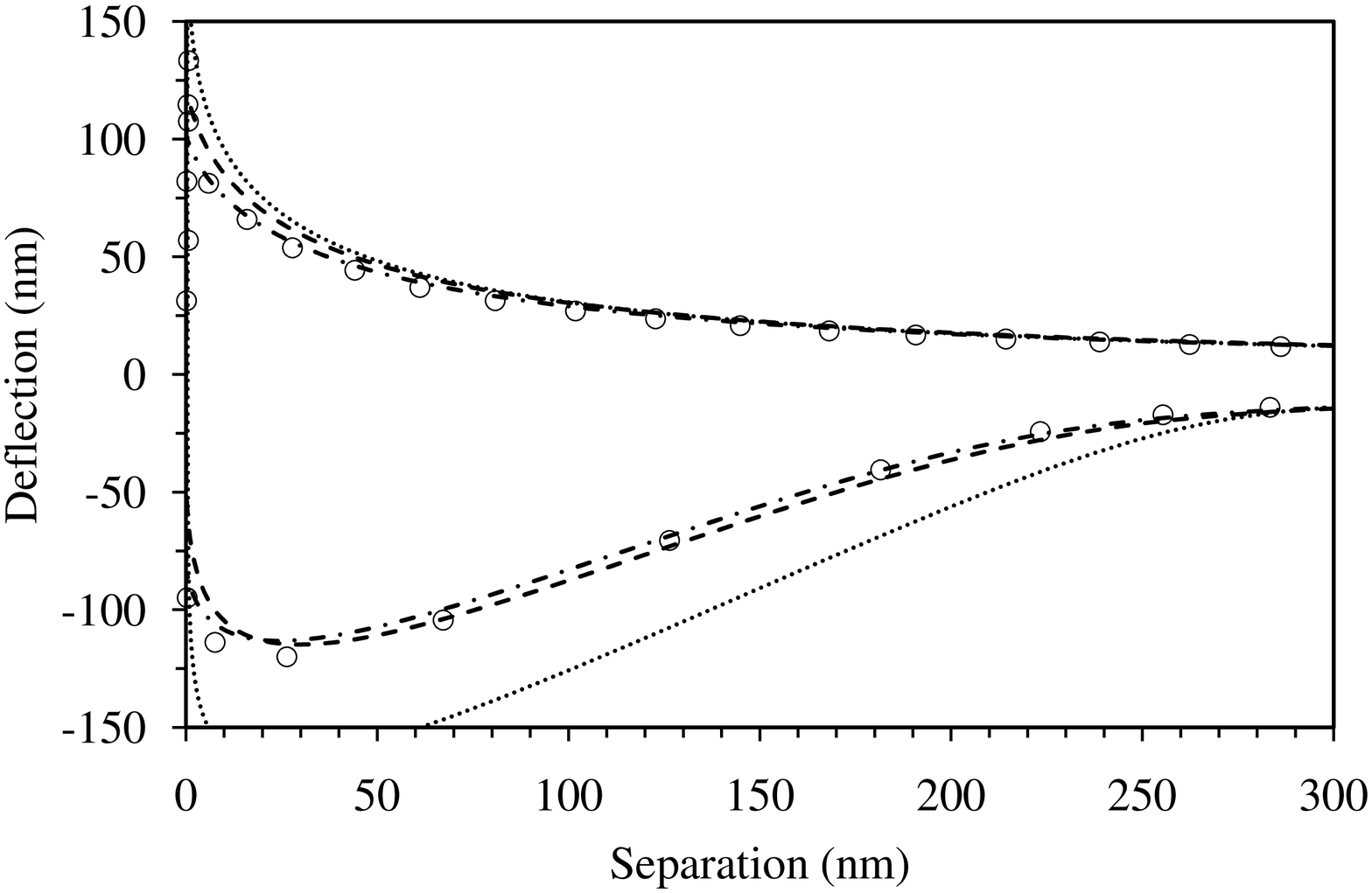} } }
\caption{\label{Fig:zvsh-SiOTS50}
Vertical cantilever deflection $z_\mathrm{c}$ versus separation $h$
at a drive rate of $\dot z_\mathrm{p} = 50\,\mu$m$\,$s$^{-1}$
for Si-OTS (see Table~\ref{tb:I}).
The effective cantilever spring constant is $k_\mathrm{eff} = 1.34\,$N/m.
The circles are the analysed experimental measurement
(every tenth point plotted),
the dotted curve is the stick calculation,
$b_\mathrm{Si}=b_\mathrm{OTS}=0$\,nm,
the dashed curve is a slip calculation,
$b_\mathrm{Si}=3$\,nm and $b_\mathrm{OTS}=2$\,nm,
both with a Hamaker constant of $A=1\times10^{-20}\,$J$\,$m$^{-2}$,
and the dash-dotted curve is a slip calculation
with $b_\mathrm{Si}=3$\,nm, $b_\mathrm{OTS}=10$\,nm,
and $A=2\times10^{-20}\,$J$\,$m$^{-2}$.
The source of the raw measured data is Ref.~\onlinecite{Zhu11b}.
}
\end{figure}

Figure \ref{Fig:zvsh-SiOTS50} shows the measured and calculated
drainage force for Si-OTS. Given the similarity in contact angles,
the OTS surface is expected to exhibit similar slip characteristics
to the DCDMS surface,
$b_\mathrm{DCDMS}=2$\,nm found above.
Although  $b_\mathrm{OTS}=2$\,nm is not a bad fit,
it can be seen that at small separations on extension,
at this velocity the measured data is better fitted with a
larger slip length, $b_\mathrm{OTS}=10$\,nm.
Both curves use $b_\mathrm{Si}=3$\,nm,
which was the value found above.
For the retract curve, both slip lengths are about equally good.
The calculations in the figure were obtained with the variable drag algorithm.
There is negligible difference for constant drag.

\begin{figure}
\centerline{
\resizebox{8.5cm}{!}{ \includegraphics*{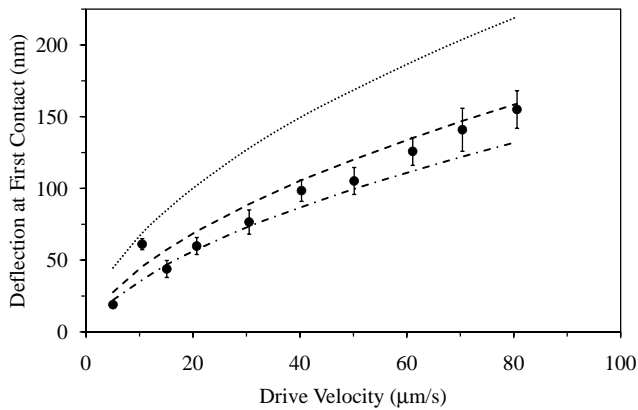} } }
\caption{\label{Fig:1stct-SiOTS}
Deflection $z_\mathrm{c}$ at first contact
versus drive velocity for Si-OTS.
The symbols, curves, and parameters are as in the preceding figure.
The bars are a crude estimate of the error arising
from the uncertainty in identifying first contact.
}
\end{figure}

For the deflection at first contact,
Fig.~\ref{Fig:1stct-SiOTS},
as expected the stick theory overestimates the repulsion
due to the drainage force.
Within the experimental scatter,
there is little to choose between a slip length of
$b_\mathrm{OTS}=2$\,nm and $b_\mathrm{OTS}=10$\,nm.
Perhaps the former is slightly better at high velocities
and the latter is slightly better at low velocities.
The Hamaker constant has almost no effect on extension (approach).

\begin{figure}
\centerline{
\resizebox{8.5cm}{!}{ \includegraphics*{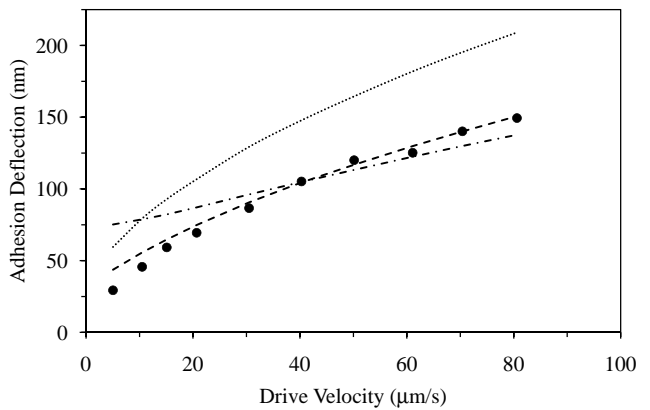} } }
\caption{\label{Fig:Adhsn-SiOTS}
Adhesion deflection $z_\mathrm{c}$ versus drive velocity for Si-OTS.
The symbols, curves, and parameters are as in the preceding figure.
The adhesive tension is $F_z = k_\mathrm{eff} z_\mathrm{c}$.
}
\end{figure}

The adhesion deflection for Si-OTS is shown in Fig.~\ref{Fig:Adhsn-SiOTS}.
As for the preceding cases,
the adhesion increases with increasing velocity,
and at each velocity it has about the same magnitude
as the deflection at first contact.
It can be seen from Fig.~\ref{Fig:Adhsn-SiOTS}
that the calculation $b_\mathrm{Si}=3$\,nm,
$b_\mathrm{OTS}=10$\,nm and $A=2\times10^{-20}\,$J$\,$m$^{-2}$
overestimates the adhesion by about 150\% at low drive velocities
and underestimates it by about 10\% at high velocities.
Decreasing the Hamaker constant to
$A=1\times10^{-20}\,$J$\,$m$^{-2}$
improves the fit at low velocities,
where it overestimates the adhesion by about 40\%,
but makes it worse at high velocities,
where it underestimates the adhesion  by about 15\%.
Overall, the slip length
$b_\mathrm{OTS}=2$\,nm
is in significantly better agreement with the measured adhesion
than is the slip length $b_\mathrm{OTS}=10$\,nm.

For the case of Si-OTS,
the deflection at first contact cannot definitively
distinguish between $b_\mathrm{OTS}=2$\,nm and $b_\mathrm{OTS}=10$\,nm.
However, based on the results for the adhesion deflection,
one can say that the slip length is much closer to
$b_\mathrm{OTS}=2$\,nm than it is to $b_\mathrm{OTS}=10$\,nm.
The former value is the same as that found for the
similar hydrophobic surface, $b_\mathrm{DCDMS}=2$\,nm.

\comment{ 
At $\dot z_\mathrm{p} = 2\,\mu$m$\,$s$^{-1}$,
the measured adhesion deflection is 29.5\,nm.
Fixing $b_\mathrm{Si}=3$\,nm,
for $b_\mathrm{OTS}=2$\,nm and $A=1\times10^{-20}\,$J$\,$m$^{-2}$
it is 43.7\,nm,
for $b_\mathrm{OTS}=10$\,nm and $A=2\times10^{-20}\,$J$\,$m$^{-2}$
it is 75.2\,nm,
and for $b_\mathrm{OTS}=10$\,nm and $A=1\times10^{-20}\,$J$\,$m$^{-2}$
it is 41.5\,nm.

At $\dot z_\mathrm{p} = 50\,\mu$m$\,$s$^{-1}$,
the measured adhesion deflection is 120.0\,nm.
Fixing $b_\mathrm{Si}=3$\,nm,
for $b_\mathrm{OTS}=2$\,nm and $A=1\times10^{-20}\,$J$\,$m$^{-2}$
it is 116.8\,nm,
for $b_\mathrm{OTS}=10$\,nm and $A=2\times10^{-20}\,$J$\,$m$^{-2}$
it is 113.2\,nm,
and for $b_\mathrm{OTS}=10$\,nm and $A=1\times10^{-20}\,$J$\,$m$^{-2}$
it is 96.0\,nm.

At $\dot z_\mathrm{p} = 80\,\mu$m$\,$s$^{-1}$,
the measured adhesion deflection is 149.4\,nm.
Fixing $b_\mathrm{Si}=3$\,nm,
for $b_\mathrm{OTS}=2$\,nm and $A=1\times10^{-20}\,$J$\,$m$^{-2}$
it is 150.7\,nm,
for $b_\mathrm{OTS}=10$\,nm and $A=2\times10^{-20}\,$J$\,$m$^{-2}$
it is 137.7\,nm,
and for $b_\mathrm{OTS}=10$\,nm and $A=1\times10^{-20}\,$J$\,$m$^{-2}$
it is 123.3\,nm.
} 

%
\section{Conclusion}
%

The non-linear algorithm for the analysis of atomic force microscopy data
has been improved and made more reliable here.
Specifically, the conversion of voltage to cantilever deflection
is now more reliable on extension out of contact.
Also the spring constant is more reliable
as it is now determined by a least squares fitting procedure.

Previously measured raw data\cite{Zhu12,Zhu11b}
was re-analyzed with the non-linear algorithm.
For silica, the slip length was found to be 3\,nm,
for dichlorodimethylsilane  it was 2\,nm,
and for octadecyltrichlorosilane it was 2\,nm.
These values are from 3--15 times smaller than those
originally obtained with the linear analysis of the same data.
\cite{Zhu12,Zhu11b}
It is of note that the two low energy surfaces have a smaller slip length
than the high energy silica surface.

Hamaker constants for the van der Waals force
were also obtained from the measured retraction data.
For Si-DOPC-Si the Hamaker constant was found to be
$1 \times 10^{-21}$\,J\,m$^{-2}$,
for DCDMS-DOPC-DCDMS it was $5 \times 10^{-20}$\,J\,m$^{-2}$,
and for Si-DOPC-OTS it was $1 \times 10^{-20}$\,J\,m$^{-2}$.
In so far as OTS is similar to DCDMS,
the fact that the asymmetric system
lies intermediate between the two symmetric systems
is physically reasonable.

\acknowledgements
The raw data used here and originally analyzed
in Refs~\onlinecite{Zhu11b,Zhu12}
were measured by Liwen Zhu
under the supervision of Chiara Neto,
and I thank Dr Zhu for providing the data.


\appendix
\section{Improvements in Atomic Force Microscopy} \label{Sec:AppA}

One of the crucial issues is the calibration of the cantilever
that is used to measure the surface forces:
using too small a value for the spring constant
causes the slip length to be overestimated
and conversely if it is too large.
Hence motivated by the need for accurate values,
one significant improvement in atomic force measurement methodology
was the \emph{in situ} measurement of
the spring constant of the cantilever.\cite{Craig01}
This was done by measuring the drainage force at large separations,
and only required values of the radius of the colloid probe
and the viscosity of the liquid, both of which can be readily measured.
This method has three advantages:
First, as an  \emph{in situ} measurement
the effective spring constant that emerges
is the one that is required for the actual measurements
that are being performed.
In particular, it automatically takes into account
the position at which the colloid
probe is glued to the cantilever
(the cantilever spring constant varies
as the cube of the distance from the base),
and also the tilt of the cantilever,
which is typically about $-11^\circ$
(the force measuring spring constant varies
as the square of the cosine of the tilt angle).
Second, the statistical error is much reduced,
being on the order of 1\% for the hydrodynamic method
compared to 10--20\% for the thermal method, for example.
Third, as has been pointed out,\cite{Attard06a}
the most common conventional method
of spring constant calibration,
the thermal method,
suffers from a mathematical error
in an early publication,\cite{Higgins06}
which has been faithfully reproduced
in the built-in software of at least one brand of atomic force microscope,
and which causes a systematic overestimate of the spring constant
of 15\%--30\%.\cite{Higgins06,Attard06a}
(The correct formulae for the thermal calibration
is given in Eq.~(35) of Ref.~\onlinecite{Attard12}.)

It is again emphasized
that an error of 10\% in a first order quantity such as the spring constant
can translate into an order of magnitude error
in a second order quantity such as the slip length.

If the spring constant calibration
can be taken as the first generation,
then the second generation of improvements
to slip length and drainage force measurement
were undertaken by the present author with colleagues Zhu and Neto
and presented in a series of papers.
\cite{Zhu11a,Zhu11b,Zhu12a,Zhu12}
Briefly these are:
\begin{itemize}
\item
a better numerical method of calculating the drainage force
that is independent of the measurement,
and a way of fitting it to the measured data that is more
sensitive to the slip length than previous methods
\item
\emph{in situ} temperature measurement,
which is important because the viscosity is sensitive to temperature,
which can vary over the course of a series of measurements
\item
a `blind test' protocol, which is used to establish the statistical accuracy
of the methods used to fit the spring constant and the slip length
\item
care to exclude particle contamination,
which creates an artefact that shows up as a very large slip length
\item
an account of thermal drift and virtual deflection
in the data analysis,
which can effect both the spring constant determination
and the slip length fit
\item
accounting for the variation in the drag force
on the cantilever with cantilever deflection,
the neglect of which causes the slip length to be overestimated
and to become cantilever- and spring constant-dependent
\end{itemize}
Of these various improvements,
perhaps the most novel is
the variation in the drag force with deflection.
This explained quantitatively many of the puzzling results
for the slip length that had been reported previously in the literature,
including the observation that it appeared to be larger for softer
cantilevers and that it appeared to depend upon the type of cantilever used.
Whether or not the non-constant cantilever drag force
is important in other measurements with the atomic force microscope
depends upon experimental details.
In general terms, reducing the drive speed
and increasing the stiffness of the cantilever
reduces the influence of the variation in drag force.

What might be called the third generation of improvements
consist primarily of taking into account
non-linear effects in the raw atomic force microscope data.\cite{Attard12}
Such non-linearity is a general feature of the atomic force microscope,
and the corrections for it have general application beyond the drainage force.
The non-linearity is illustrated in Fig.~\ref{Fig:VvsZp},
where the measured photo-diode voltage is plotted
against the piezo-drive position approaching and in contact.
The conventional analysis of atomic microscope force data
begins by taking the slope of the line in contact, V/nm,
the so-called constant compliance factor,
and using it directly to convert a change in voltage
to a change in cantilever deflection, and hence,
using the spring constant, to a change in force.
This is how forces are measured with the atomic force microscope.
However as can be seen in  Fig.~\ref{Fig:VvsZp},
the contact region does not have constant slope,
and whatever value of the slope is chosen as the calibration factor
has a quantitative effect on the data analysis
that gives the measured forces.
The effect is quite dramatic in the contact region,
as can be seen in the inset.
However,  even though the effect seems small for the pre-contact drainage force
on the scale of the figure,
it can change by more than an order of magnitude the fitted slip length.
\cite{Attard12}

In Ref.~\onlinecite{Attard12} two sources of non-linearity were analysed:
non-linear photo-diode response and non-linear cantilever deflection.
It was concluded that in typical experiments the latter was negligible,
and that it was the photo-diode that was responsible for the non-linearity
evident in Fig.~\ref{Fig:VvsZp}.
(Most likely, it is the elliptical cross-section
and Gaussean intensity distribution of the light beam
that directly causes the non-linearity.)
The spreadsheet algorithm that was developed in Ref.~\onlinecite{Attard12}
to analyze surface forces with a non-linear contact region
has a number of features:
\begin{itemize}
\item
it removes the ambiguity in the calibration factor
by using a polynomial fit
for the non-linear conversion of the measured voltage to cantilever deflection
\item
relying upon earlier analysis,\cite{Attard98,Attard99}
it takes into account cantilever tilt, friction, and torque
to distinguish between the intrinsic and the effective cantilever
spring constant, and to obtain the actual pre-contact force
from the raw voltage
\item
it automates the identification of the zero of separation,
which avoids human intervention or bias,
and which can be particularly problematic when the linear analysis
gives an unphysical contact region
\item
it allows the simultaneous analysis of both the extension and
the retraction data so that the drainage adhesion
can be used as an additional constraint on
the determination of the slip length
\end{itemize}

This third generation analysis
was applied to the same raw data
as was previously analysed with the second generation technique,
namely the drainage force between silica surfaces
in di-n-octylphthalate (DOPC).
Whereas the second generation linear analysis
had given the slip length as
$11\pm 2.5\,$nm,\cite{Zhu12}
with the non-linear analysis it was found to be
$3\pm 1\,$nm.\cite{Attard12}
In addition, the linear analysis had shown that the slip length
decreased with increasing shear rate,
whereas the non-linear analysis had shown that it was constant.

The reasons why the linear analysis
had these particular quantitative and qualitative failings
was discussed in detail in the conclusion of Ref.~\onlinecite{Attard12}.
Briefly, due to the fact that the compliance curve, Fig.~\ref{Fig:VvsZp},
is concave down,
the compliance slope taken from the tangent at first contact
is an underestimate of the  factor required in the non-contact region.
This means that the effective spring constant fitted at large separations
is too small
($k_\mathrm{eff} = 1.5\,$N/m,\cite{Zhu12}
compared to $k_\mathrm{eff} = 1.68\,$N/m\cite{Attard12}
obtained with the non-linear analysis),
and consequently
the slip length required to fit the data at intermediate separations
is too large.
This overcorrects the problem  closer to contact,
where the calibration factor that is used
is closer to the real one.
Since the shear rate increases with decreasing separation,
the reduced slip length required at small separations
was interpreted as being due to shear rate dependence.\cite{Zhu12a}
In the non-linear analysis of Ref.~\onlinecite{Attard12},
a single slip length gave a good fit over all separations
(down to about 1~nm) and for all drive velocities.
As well, the retraction data,
which was not originally analysed \cite{Zhu12},
was also well fitted using the non-linear analysis,
over almost the whole separation regime and at all drive velocities
using the same slip length.

%
\section{Non-Linear Data Analysis Algorithm for a Spread-Sheet}
\label{Sec:AppB}
%

This section sets out in order the steps
for the spread-sheet algorithm
for the non-linear analysis of raw data
measured with the atomic force microscope.
The section is written in the form of a recipe,
and only limited justification and explanation is given.
The formula, with small modifications,
are derived in Ref.~\onlinecite{Attard12},
with the linear cantilever coefficients
originally derived in Refs.~\onlinecite{Attard98,Attard99}.

Here it is assumed that the raw extension
and retraction data are separated.
These quantities are denoted  `ext' and `ret' respectively,
and each is analysed independently except as noted below.
Further `c' refers to contact, `nc' refers to non-contact,
and `b' refers to base line.

In the atomic force microscope modeled here,
the piezo-drive is connected to the cantilever holder
and is above the substrate.
In extension the velocity of the piezo-drive is negative,
$\dot z_\mathrm{p} < 0$,
and in contact the movement of the contact point
is equal and opposite to that of the piezo-drive,
$\Delta z_\mathrm{c} = -\Delta z_\mathrm{p}$.
In other models of the atomic force microscope,
it is the substrate that is connected to the  piezo-drive,
so that $\dot z_\mathrm{p} > 0$ on extension
and in contact $\Delta z_\mathrm{c} = \Delta z_\mathrm{p}$.
These other models may be analysed with the formulae given in this paper
by negating the piezo-drive position in the raw  data,
$z_\mathrm{p} \Rightarrow -z_\mathrm{p}$.

\subsection{Base Line}

Select the base line region, where the raw voltage is a linear function
of the drive distance.
Do not include the initial extension data or the final retraction data
where the piezo-drive is not moving at constant velocity.
Do not include (or include as little as possible)
of the region where the data is clearly curved due to the drainage force.

Independently fit two straight lines
to the measured voltages in the  base line regions:
\begin{equation}
V_\mathrm{b}^\mathrm{ext}(z_\mathrm{p})
=
V_\mathrm{b}^\mathrm{ext}
+
V_\mathrm{b}^\mathrm{ext'}[z_\mathrm{p}-z_\mathrm{pb}^\mathrm{ext}] ,
\end{equation}
and
\begin{equation}
V_\mathrm{b}^\mathrm{ret}(z_\mathrm{p})
=
V_\mathrm{b}^\mathrm{ret}
+
V_\mathrm{b}^\mathrm{ret'}[z_\mathrm{p}-z_\mathrm{pb}^\mathrm{ret}] .
\end{equation}
These include the effects of virtual deflection, thermal drift,
drag, and drainage force at long range.
Different spread sheet programs fit straight lines in different formats.
It is assumed that the user can convert the fitted form to the above format,
with $z_\mathrm{pb}$ chosen to be in the middle of the selected base line region
by the user.
Experience indicates that best results are obtained with
$z_\mathrm{pb}^\mathrm{ext} \approx z_\mathrm{pb}^\mathrm{ret}$.

\subsection{Force Asymptote}

Calculate the linear asymptote
to the drainage force in the base line region:
\begin{equation}
F_\mathrm{b}^\mathrm{ext}(z_\mathrm{p})
=
F_\mathrm{b}^\mathrm{ext}
+
F_\mathrm{b}^\mathrm{ext'}[z_\mathrm{p}-z_\mathrm{pb}^\mathrm{ext}] ,
\end{equation}
and
\begin{equation}
F_\mathrm{b}^\mathrm{ret}(z_\mathrm{p})
=
F_\mathrm{b}^\mathrm{ret}
+
F_\mathrm{b}^\mathrm{ret'}[z_\mathrm{p}-z_\mathrm{pb}^\mathrm{ret}] .
\end{equation}
This requires that the separation be known,
$h = z_\mathrm{p} + z_\mathrm{c} + z_\mathrm{0}^\mathrm{ext/ret} $.
The constants $z_\mathrm{0}^\mathrm{ext}$ and $z_\mathrm{0}^\mathrm{ret} $
are obtained in \S\ref{Sec:z0} below.
The vertical deflection of the contact point  is
$ z_\mathrm{c} = F_z / k_\mathrm{eff}$.
With $C \equiv - 6 \pi \eta R^2$,
one can set up the iteration
\begin{equation}
F_z^{(0)} = \frac{C \dot z_\mathrm{p}}{z_0+z_\mathrm{p}} ,
\mbox{ and }
F_z^{(n)} =
\frac{C \dot z_\mathrm{p} - k_\mathrm{eff}^{-1} [ F_z^{(n-1)}]^2
}{
z_0+z_\mathrm{p} + k_\mathrm{eff}^{-1} F_z^{(n-1)} }.
\end{equation}
Generally only zero or one iterations are required.
One can recognize here
$h^{(n)} = z_0+z_\mathrm{p} + k_\mathrm{eff}^{-1} F_z^{(n)}$.
For the base line, choose $z_\mathrm{p} = z_\mathrm{pb}$
and this iterative formula gives the base line constants
$F_\mathrm{b}$ and  $F_\mathrm{b}' = -F_\mathrm{b}/h$.
Obviously one uses either $\dot z_\mathrm{p}^\mathrm{ext}$
or $\dot z_\mathrm{p}^\mathrm{ret}$
for extension or retraction, respectively.

Its worth mentioning that the algorithm given in \S\ref{Sec:AlgoDrain}
below was also used for the extension branch.
It is slightly more robust than the present one,
but it makes no difference to the fitted spring constant.
The present algorithm has the advantage that it does not need
to calculate the retraction curve from contact.

In the above,
$\eta$ is the viscosity,
$R$ is the radius of the colloid probe,
and $k_\mathrm{eff}$ is the effective cantilever spring constant.
The method of determining the  spring constant is given below.
Fortunately, the results are not very sensitive to this value,
and so in the first instance any estimate can be used,
and this can later be replaced by the accurate value.

\subsection{Contact Fit}

Perform non-linear fits to the measured voltage in the contact regions:
\begin{equation}
z_\mathrm{pc}^\mathrm{ext}(V)
=
a^\mathrm{ext} + b^\mathrm{ext} V + c^\mathrm{ext} V^2
+ d^\mathrm{ext} V^3 + \ldots ,
\end{equation}
and
\begin{equation}
z_\mathrm{pc}^\mathrm{ret}(V)
=
a^\mathrm{ret} + b^\mathrm{ret} V + c^\mathrm{ret} V^2
+ d^\mathrm{ret} V^3 + \ldots .
\end{equation}
Note the distinction between $z_\mathrm{pc}$,
which is the piezo-drive position at a given voltage
when the probe is in contact with the substrate,
and $z_\mathrm{c}$,
which is the vertical deflection of the contact position (apex) of the probe.
In practice four terms were found adequate.
One should probably avoid increasing the number of terms.

It is important to appreciate that the
extension fit only covers voltages strictly greater than the voltages
measured in the non-contact extension region.
In contrast, the retraction fit covers almost all measured voltages,
both extension and retraction, contact and non-contact.
(The exception is the voltages around the minimum of the retract curve,
which is not a large extrapolation anyway.)
It has been found, for example,
that the slope of the contact voltage
evaluated at the extension base line voltage
is unreliable when extrapolated from the extension fit,
but is reliable when interpolated from the retraction fit.
(Reliability here was judged by whether or not it remained more or less
constant over a long series of measurements.)
The difference was around 5\% in a typical case.
For this reason the retraction fit will primarily be used
for the non-contact data in what follows.
This is one of two improvements on the algorithm
given in Ref.~\onlinecite{Attard12}.

It is not appropriate to use the retraction fit for the extension data
in contact.
This is because friction plays an opposite role in the two cases,
and also because in contact the extension fit is strictly an interpolation
for extension in contact
(except possibly near the ends).
Hence as explained below a switch is used
to apply the two different fits to the extension data
in the two regimes.
(The retraction fit can be used for all the retraction data.)
Such a switch is only necessary if one wants to look at the topography
in contact.
It does require a fairly accurate determination of the point of first contact.

\subsection{Compliance Slope}

One eventually has to convert the base line voltages
to a deflection angle,
and for this the compliance factor derived from
the gradient of the contact fit
evaluated at the base line voltage is required.
The constant compliance gradients for the base line
are defined as
\begin{eqnarray}
\beta_\mathrm{cb}^\mathrm{ext}
& \equiv &
\left. \frac{\mathrm{d} V_\mathrm{c}^\mathrm{ext}(z_\mathrm{p})
}{\mathrm{d}z_\mathrm{p} } \right|_{V_\mathrm{b}^\mathrm{ext}}
\nonumber \\ & = &
\left. \frac{\mathrm{d} V
}{\mathrm{d}\theta } \right|_{V_\mathrm{b}^\mathrm{ext}}
\left. \frac{\mathrm{d} \theta }{\mathrm{d}z_\mathrm{c} } \right|_\mathrm{ext}
\left. \frac{\mathrm{d} z_\mathrm{c} }{\mathrm{d}z_\mathrm{p} }
\right|_\mathrm{c}
\left. \frac{\mathrm{d} \theta }{\mathrm{d}z_\mathrm{c} } \right|_\mathrm{ret}
\left. \frac{\mathrm{d}z_\mathrm{c} }{\mathrm{d} \theta } \right|_\mathrm{ret}
\nonumber \\ & = &
\left. \frac{\mathrm{d} V_\mathrm{c}^\mathrm{ret}(z_\mathrm{p})
}{\mathrm{d}z_\mathrm{p} } \right|_{V_\mathrm{b}^\mathrm{ext}}
\frac{\alpha^\mathrm{ext}}{\alpha^\mathrm{ret}}
\nonumber \\ & = &
\frac{ \alpha^\mathrm{ext} / \alpha^\mathrm{ret}
}{
b^\mathrm{ret}  + 2 c^\mathrm{ret} V_\mathrm{b}^\mathrm{ext}
+ 3 d^\mathrm{ret} [V_\mathrm{b}^\mathrm{ext}]^2  + \ldots } .
\end{eqnarray}
Here $\theta$ is the angle of deflection of the cantilever.
The formula for
$\alpha \equiv \mathrm{d} \theta / \mathrm{d}z_\mathrm{c} $
is given in \S\ref{Sec:CantChar}.

Notice that only the retraction fit is used for this.
In order to convert from the extension slope to the retraction slope
one has to account for difference between the rates of change of angle
on extension and retraction in contact,
which is the reason for the ratio in the numerator.

The retraction gradient is
\begin{eqnarray}
\beta_\mathrm{cb}^\mathrm{ret}
& \equiv &
\left.
\frac{\mathrm{d} V_\mathrm{c}^\mathrm{ret}(z_\mathrm{p})
}{\mathrm{d}z_\mathrm{p} }
\right|_{V_\mathrm{b}^\mathrm{ret}}
\nonumber \\ & = &
\frac{1}{ b^\mathrm{ret}  + 2 c^\mathrm{ret} V_\mathrm{b}^\mathrm{ret}
+ 3 d^\mathrm{ret} [V_\mathrm{b}^\mathrm{ret}]^2  + \ldots } .
\end{eqnarray}

These slopes are dominated by the change in angle of the cantilever
due to the surface forces in contact.
However, they also contain a contribution from the base line slope,
which should really be removed.
One could define
\begin{equation}
\beta_\mathrm{cbf}^\mathrm{ext}
\equiv
\beta_\mathrm{cb}^\mathrm{ext}
- V_\mathrm{b}^\mathrm{ext'} -
\beta_\mathrm{cb}^\mathrm{ext}
k_\mathrm{eff}^{-1} F_\mathrm{b}^\mathrm{ext'} ,
\end{equation}
and similarly for $\beta_\mathrm{cbf}^\mathrm{ret}$.
(Note that in contact, $\mathrm{d}z_\mathrm{c}  =- \mathrm{d} z_\mathrm{p} $.)
These corrections are generally negligible.
In practice this correction was implemented with $F_\mathrm{b}'=0$
(to avoid circularity)
even though the change was typically only 0.05\%.

For completeness,
the rate of change of voltage with vertical contact position
out of contact is
\begin{eqnarray}
\lambda(V)
& \equiv &
\frac{\mathrm{d} V }{\mathrm{d}z_\mathrm{c} }
 =
\frac{\mathrm{d} V }{\mathrm{d}\theta}
\left. \frac{\mathrm{d} \theta }{\mathrm{d}z_\mathrm{c} } \right|_\mathrm{nc}
\nonumber \\ & = &
\left. \frac{\mathrm{d} V_\mathrm{c}^\mathrm{ret}(z_\mathrm{p})
}{\mathrm{d}z_\mathrm{p} } \right|_\mathrm{ret}
\left. \frac{\mathrm{d} z_\mathrm{p} }{\mathrm{d}z_\mathrm{c} }
\right|_\mathrm{c}
\left. \frac{\mathrm{d}z_\mathrm{c} }{\mathrm{d} \theta } \right|_\mathrm{ret}
\left. \frac{\mathrm{d} \theta }{\mathrm{d}z_\mathrm{c} } \right|_\mathrm{nc}
\nonumber \\ & = &
\frac{-\alpha/\alpha^\mathrm{ret} }
{ b^\mathrm{ret}  + 2 c^\mathrm{ret} V + 3 d^\mathrm{ret} V^2  + \ldots } .
\end{eqnarray}
This holds for both extension and retraction out of contact.

\subsection{Contact Position}

The initial estimate of the contact position
is where the contact voltage equals the constant part of the
base line voltage,
\begin{equation}
z_\mathrm{pcb}^\mathrm{ext}
=
a^\mathrm{ret} + b^\mathrm{ret}  V_\mathrm{b}^\mathrm{ext}
+ c^\mathrm{ret}  [ V_\mathrm{b}^\mathrm{ext}]^2 + \ldots
\end{equation}
and
\begin{equation}
z_\mathrm{pcb}^\mathrm{ret}
=
a^\mathrm{ret} + b^\mathrm{ret}  V_\mathrm{b}^\mathrm{ret}
+ c^\mathrm{ret}  [ V_\mathrm{b}^\mathrm{ret}]^2 + \ldots
\end{equation}
Again the retraction fit is used for both,
but they are evaluated at the respective base line voltages.

Now this is corrected so that
the fitted voltage equals the actual linear base line voltage at that position,
$z_\mathrm{pcb}^\mathrm{ext,*}
= z_\mathrm{pc}^\mathrm{ext}
(V_\mathrm{b}^\mathrm{ext}(z_\mathrm{pcb}^\mathrm{ext,*}))$.
A Taylor expansion yields
\begin{eqnarray}
z_\mathrm{pcb}^\mathrm{ext,*} - z_\mathrm{pcb}^\mathrm{ext}
& = &
\frac{\mathrm{d} z_\mathrm{p}}{\mathrm{d} V_\mathrm{c}^\mathrm{ext}}
\left[ V_\mathrm{b}^\mathrm{ext}(z_\mathrm{pcb}^\mathrm{ext,*})
- V_\mathrm{b}^\mathrm{ext}  \right]
\nonumber \\ & = &
(\beta_\mathrm{cb}^\mathrm{ext})^{-1}
V_\mathrm{b}^\mathrm{ext'}
\left[ z_\mathrm{pcb}^\mathrm{ext,*} - z_\mathrm{pb}^\mathrm{ext}  \right] ,
\end{eqnarray}
or
\begin{equation}
z_\mathrm{pcb}^\mathrm{ext,*}
=
\frac{ \beta_\mathrm{cb}^\mathrm{ext} z_\mathrm{pcb}^\mathrm{ext}
- V_\mathrm{b}^\mathrm{ext'} z_\mathrm{pb}^\mathrm{ext}
}{\beta_\mathrm{cb}^\mathrm{ext} - V_\mathrm{b}^\mathrm{ext'} } .
\end{equation}
One has an analogous result for retraction,
\begin{equation}
z_\mathrm{pcb}^\mathrm{ret,*}
=
\frac{ \beta_\mathrm{cb}^\mathrm{ret} z_\mathrm{pcb}^\mathrm{ret}
- V_\mathrm{b}^\mathrm{ret'} z_\mathrm{pb}^\mathrm{ret}
}{\beta_\mathrm{cb}^\mathrm{ret} - V_\mathrm{b}^\mathrm{ret'} } .
\end{equation}

In so far as the compliance slope is generally much greater
than the base line slope,
the difference between $z_\mathrm{pcb}^\mathrm{ext,*}$
and $z_\mathrm{pcb}^\mathrm{ext}$ is generally small,
1.5$\,$nm in a typical case.
This is negligible at large separations
but is important at small, particularly if one wants
the topography in contact.

\subsection{Zero of Separation} \label{Sec:z0}

The deflection of the contact position due to the force
in the base line region  is for extension,
\begin{eqnarray}
z_\mathrm{cb}^\mathrm{ext}(z_\mathrm{p})
& = &
k_\mathrm{eff}^{-1} F_\mathrm{b}^\mathrm{ext}(z_\mathrm{p})
\nonumber \\ &=&
k_\mathrm{eff}^{-1}
\left\{ F_\mathrm{b}^\mathrm{ext}
+
F_\mathrm{b}^\mathrm{ext'}
\left[ z_\mathrm{p} - z_\mathrm{pb}^\mathrm{ext}\right] \right\} ,
\end{eqnarray}
and for retraction,
\begin{eqnarray}
z_\mathrm{cb}^\mathrm{ret}(z_\mathrm{p})
& = &
k_\mathrm{eff}^{-1} F_\mathrm{b}^\mathrm{ret}(z_\mathrm{p})
\nonumber \\ &=&
k_\mathrm{eff}^{-1}
\left\{ F_\mathrm{b}^\mathrm{ret}
+
F_\mathrm{b}^\mathrm{ret'}
\left[ z_\mathrm{p} - z_\mathrm{pb}^\mathrm{ret}\right] \right\} .
\end{eqnarray}

These linear results are defined as holding until contact,
even though the actual measured forces depart significantly
from the base line.
Recall that the contact point is defined
as the intersection of the extrapolated base line
and the extrapolated contact curve.

The separation is in general $h=z_\mathrm{p}+z_\mathrm{c}+z_0$,
with the constant $z_0$ to be now determined.
At contact, $z_\mathrm{p}= z_\mathrm{pcb}^\mathrm{*}$,
one must have $h=0$.
Hence
\begin{eqnarray}
z_0^\mathrm{ext}
& = &
-z_\mathrm{pcb}^\mathrm{ext,*}
-z_\mathrm{cb}^\mathrm{ext}(z_\mathrm{pcb}^\mathrm{ext,*}) ,
\end{eqnarray}
and at $z_\mathrm{pcb}^\mathrm{ret,*}$,
\begin{eqnarray}
z_0^\mathrm{ret}
& = &
-z_\mathrm{pcb}^\mathrm{ret,*}
-z_\mathrm{cb}^\mathrm{ret}(z_\mathrm{pcb}^\mathrm{ret,*}) .
\end{eqnarray}
These give the zero of separation.

It will be noted that there is no direct human intervention
in the choice of the zero of separation.
These constants emerge automatically from the fits
to the base line and contact voltages.

\subsection{Base Line Deflection Angle}

The base line angle for extension is
\begin{eqnarray}
\theta_\mathrm{b}^\mathrm{ext}(z_\mathrm{p})
& = &
\theta_\mathrm{b}^\mathrm{ext}
+
\left. \frac{\mathrm{d}\theta}{\mathrm{d}z_\mathrm{c}} \right|_\mathrm{nc}
\left. \frac{\mathrm{d}z_\mathrm{c}}{\mathrm{d}V} \right|_\mathrm{nc}
V_\mathrm{b}^\mathrm{ext'}
\left[ z_\mathrm{p} - z_\mathrm{pb}^\mathrm{ext}\right]
\nonumber \\ & = &
\theta_\mathrm{b}^\mathrm{ext}
+
\frac{\alpha}{ \lambda(V_\mathrm{b}^\mathrm{ext}) }
V_\mathrm{b}^\mathrm{ext'}
\left[ z_\mathrm{p} - z_\mathrm{pb}^\mathrm{ext}\right]
\nonumber \\ & = &
\theta_\mathrm{b}^\mathrm{ext}
-
\frac{\alpha^\mathrm{ext} }{ \beta_\mathrm{cb}^\mathrm{ext} }
V_\mathrm{b}^\mathrm{ext'}
\left[ z_\mathrm{p} - z_\mathrm{pb}^\mathrm{ext}\right] .
\end{eqnarray}
The two final equalities follow either by direct substitution,
or else by evaluating the derivatives in the first equality
in contact on extension.
Similarly, that for retraction is
\begin{eqnarray}
\theta_\mathrm{b}^\mathrm{ret}(z_\mathrm{p})
& = &
\theta_\mathrm{b}^\mathrm{ret}
+
\frac{\alpha}{ \lambda(V_\mathrm{b}^\mathrm{ret}) }
V_\mathrm{b}^\mathrm{ret'}
\left[ z_\mathrm{p} - z_\mathrm{pb}^\mathrm{ret}\right]
\nonumber \\ & = &
\theta_\mathrm{b}^\mathrm{ret}
-
\frac{\alpha^\mathrm{ret} }{ \beta_\mathrm{cb}^\mathrm{ret} }
V_\mathrm{b}^\mathrm{ret'}
\left[ z_\mathrm{p} - z_\mathrm{pb}^\mathrm{ret}\right] .
\end{eqnarray}

There are two undetermined constant angles here:
$\theta_\mathrm{b}^\mathrm{ext}$
and $\theta_\mathrm{b}^\mathrm{ret}$.
Since the difference in angles is linearly proportional
to the difference in base line voltages,
a symmetric choice is
\begin{equation}
\theta_\mathrm{b}^\mathrm{ret}
=
-\theta_\mathrm{b}^\mathrm{ext}
=
\frac{\alpha}{2 \lambda(V_\mathrm{b}) }
\left[ V_\mathrm{b}^\mathrm{ret} - V_\mathrm{b}^\mathrm{ext} \right] ,
\end{equation}
with
$V_\mathrm{b} \equiv
[V_\mathrm{b}^\mathrm{ret} + V_\mathrm{b}^\mathrm{ext} ] /2$.
In fact, these two constants drop out of the final formula below
and so the actual values are immaterial.

The deflection angle due to the base line force for retraction is
\begin{eqnarray}
\theta_\mathrm{bf}^\mathrm{ret}(z_\mathrm{p})
& = &
\frac{\alpha}{k_\mathrm{eff}} \left\{
F_\mathrm{b}^\mathrm{ret}
+
F_\mathrm{b}^\mathrm{ret'}
\left[ z_\mathrm{p}- z_\mathrm{pb}^\mathrm{ret} \right] \right\},
\end{eqnarray}
and that for extension is
\begin{equation}
\theta_\mathrm{bf}^\mathrm{ext}(z_\mathrm{p})
=
\frac{\alpha}{k_\mathrm{eff}} \left\{
F_\mathrm{b}^\mathrm{ext}
+ F_\mathrm{b}^\mathrm{ext'}
\left[ z_\mathrm{p}- z_\mathrm{pb}^\mathrm{ext} \right] \right\} .
\end{equation}
For convenience below the constant parts of these may be defined as
$ \theta_\mathrm{bf}^\mathrm{ret} \equiv
 \alpha F_\mathrm{b}^\mathrm{ret} /k_\mathrm{eff}$
and
$ \theta_\mathrm{bf}^\mathrm{ext} \equiv
 \alpha F_\mathrm{b}^\mathrm{ext} /k_\mathrm{eff}$.

\subsection{Deflection Angle}

All the formulae above have been for constants,
or linear base line fits,
or polynomial contact fits.
Now is given the formulae that convert the measured raw voltage
into a deflection angle of the cantilever.
Obviously the formulae will invoke
the fits and constants given above.

In the atomic force microscope
one has a set of data pairs $\{z_\mathrm{p},V\}$,
separately for extension and for retraction.
The aim is to convert each data pair to separation and force.
Since the latter two are linear functions of the deflection angle,
and it is the components of the angle that are linearly additive
(even though the voltage is a non-linear function of the angle),
the  formula are obtained for the deflection angle.

The deflection angle on the retract contact curve is
\begin{equation}
\theta_\mathrm{c}^\mathrm{ret}(V)
=
\theta_\mathrm{c}^\mathrm{ret,*}
-
\alpha^\mathrm{ret}
\left[ z_\mathrm{pc}^\mathrm{ret}(V) - z_\mathrm{pcb}^\mathrm{ret,*} \right] .
\end{equation}
The constant $\theta_\mathrm{c}^\mathrm{ret,*}$
is the deflection angle at the intersection
of the contact curve and the linear base line,
and must be the same for both
(since by definition they have the same voltage at this point),
\begin{eqnarray}
\theta_\mathrm{c}^\mathrm{ret,*} & \equiv &
\theta_\mathrm{c}^\mathrm{ret}(V^\mathrm{ret,*})
=
\theta_\mathrm{bf}^\mathrm{ret}(V^\mathrm{ret,*})
\nonumber \\ & = &
\frac{\alpha}{k_\mathrm{eff}}
\left\{ F_\mathrm{b}^\mathrm{ret}
+
F_\mathrm{b}^\mathrm{ret'}
\left[ z_\mathrm{pcb}^\mathrm{ret,*}
- z_\mathrm{pb}^\mathrm{ret}\right] \right\} .
\end{eqnarray}

The total angle  on the retract contact curve is
\begin{equation}
\theta_\mathrm{tot,c}^\mathrm{ret}(V)
=
\theta_0 + \theta_\mathrm{c}^\mathrm{ret}(V)
+
\theta_\mathrm{b}^\mathrm{ret}(z_\mathrm{pc}^\mathrm{ret}(V))
-
\theta_\mathrm{bf}^\mathrm{ret}(z_\mathrm{pc}^\mathrm{ret}(V)) .
\end{equation}
The constant tilt angle is $\theta_0 $.
(All angles are in radians; the tilt angle is negative.)
Everything else on the right hand side was defined above.

The total angle at a given voltage and a given position
on extension is
the sum of the tilt angle,
the deflection angle,
and the base line angle
less the base line deflection angle,
\begin{eqnarray}
\theta^\mathrm{ext}_\mathrm{tot}(V,z_\mathrm{p})
=
\theta_0 + \theta^\mathrm{ext}(V,z_\mathrm{p})
+ \theta_\mathrm{b}^\mathrm{ext}(z_\mathrm{p})
- \theta_\mathrm{bf}^\mathrm{ext}(z_\mathrm{p}) .
\end{eqnarray}
Since there is a one to one correspondence between
the photodiode voltage and the total cantilever angle,
this must equal the total angle in contact on retraction
at the same voltage,
$ \theta^\mathrm{ext}_\mathrm{tot}(V,z_\mathrm{p})
= \theta_\mathrm{tot,c}^\mathrm{ret}(V)$.
Hence the deflection angle at a given voltage and a given position
on extension is
\begin{eqnarray}
\theta^\mathrm{ext}(V,z_\mathrm{p})
& = &
\theta_\mathrm{tot,c}^\mathrm{ret}(V)
- \theta_0
- \theta_\mathrm{b}^\mathrm{ext}(z_\mathrm{p})
+ \theta_\mathrm{bf}^\mathrm{ext}(z_\mathrm{p})
\nonumber \\ & = &
\theta_\mathrm{c}^\mathrm{ret}(V)
+
\theta_\mathrm{b}^\mathrm{ret}(z_\mathrm{pc}^\mathrm{ret}(V))
-
\theta_\mathrm{bf}^\mathrm{ret}(z_\mathrm{pc}^\mathrm{ret}(V))
\nonumber \\ && \mbox{ }
- \theta_\mathrm{b}^\mathrm{ext}(z_\mathrm{p})
+ \theta_\mathrm{bf}^\mathrm{ext}(z_\mathrm{p}).
\end{eqnarray}
Since
$\theta^\mathrm{ext}(V_\mathrm{b}^\mathrm{ext},z_\mathrm{pb}^\mathrm{ext})
= \theta_\mathrm{bf}^\mathrm{ext}$,
which was give above,
the constant terms may be collected and this expression may be rearranged as
\begin{eqnarray}
\lefteqn{
\theta^\mathrm{ext}(V,z_\mathrm{p})
 =}
\\ \nonumber & &
- \alpha^\mathrm{ret}
\left[ z_\mathrm{pc}^\mathrm{ret}(V) - z_\mathrm{pcb}^\mathrm{ext}  \right]
- \frac{ \alpha^\mathrm{ret} }{  \beta_\mathrm{cb}^\mathrm{ret} }
V_\mathrm{b}^\mathrm{ret'}
\left[ z_\mathrm{pc}^\mathrm{ret}(V) - z_\mathrm{pcb}^\mathrm{ext}  \right]
\nonumber \\ && \mbox{ }
-
\frac{\alpha}{k_\mathrm{eff}} F_\mathrm{b}^\mathrm{ret'}
\left[ z_\mathrm{pc}^\mathrm{ret}(V) - z_\mathrm{pcb}^\mathrm{ext}  \right]
+ \frac{ \alpha^\mathrm{ext} }{  \beta_\mathrm{cb}^\mathrm{ext} }
V_\mathrm{b}^\mathrm{ext'}
\left[ z_\mathrm{p} - z_\mathrm{pb}^\mathrm{ext} \right]
\nonumber \\ \nonumber && \mbox{ }
+ \frac{\alpha}{k_\mathrm{eff}} F_\mathrm{b}^\mathrm{ext'}
\left[ z_\mathrm{p}- z_\mathrm{pb}^\mathrm{ext} \right]
+
\theta_\mathrm{bf}^\mathrm{ext} ,
\end{eqnarray}
since
$z_\mathrm{pcb}^\mathrm{ext} \equiv
z_\mathrm{pc}^\mathrm{ret}(V_\mathrm{b}^\mathrm{ext})$.
This result hold on extension out of contact.
(See the end of this section for the contact formula.)

The analogous result for retraction is
\begin{eqnarray}
\lefteqn{
\theta^\mathrm{ret}(V,z_\mathrm{p})
=
} \\  \nonumber &&
- \alpha^\mathrm{ret}
\left[ z_\mathrm{pc}^\mathrm{ret}(V) - z_\mathrm{pcb}^\mathrm{ret}  \right]
- \frac{ \alpha^\mathrm{ret} }{  \beta_\mathrm{cb}^\mathrm{ret} }
V_\mathrm{b}^\mathrm{ret'}
\left[ z_\mathrm{pc}^\mathrm{ret}(V) - z_\mathrm{pcb}^\mathrm{ret}  \right]
\nonumber \\ && \mbox{ }
-
\frac{\alpha}{k_\mathrm{eff}} F_\mathrm{b}^\mathrm{ret'}
\left[ z_\mathrm{pc}^\mathrm{ret}(V) - z_\mathrm{pcb}^\mathrm{ret}  \right]
+ \frac{ \alpha^\mathrm{ret} }{  \beta_\mathrm{cb}^\mathrm{ret} }
V_\mathrm{b}^\mathrm{ret'}
\left[ z_\mathrm{p} - z_\mathrm{pb}^\mathrm{ret} \right]
\nonumber \\ \nonumber && \mbox{ }
+ \frac{\alpha}{k_\mathrm{eff}} F_\mathrm{b}^\mathrm{ret'}
\left[ z_\mathrm{p}- z_\mathrm{pb}^\mathrm{ret} \right]
+ \theta_\mathrm{bf}^\mathrm{ret} .
\end{eqnarray}

From these deflection angles,
the vertical deflection of the contact position
can be obtained, $z_\mathrm{c}  = \alpha^{-1}  \theta$,
with the constant given in \S\ref{Sec:CantChar}.
From the deflection and the constants given above,
the separation  can be obtained, $h = z_\mathrm{p} + z_\mathrm{c} + z_0$.

As mentioned above,
it is most accurate to use the retraction contact voltage fit
to obtain the deflection angle for non-contact extension,
because only interpolation is required.
In the non-contact regime, friction has no influence,
but in contact it does, and this approach messes up the angle
for extension in contact.
In the usual case there is no need for data in contact.
However for completeness, the formula to be used for extension in contact
is as above with `ret' change to `ext',
\begin{eqnarray}
\lefteqn{
\theta^\mathrm{ext}(V,z_\mathrm{p})
=
} \\ \nonumber &&
- \alpha^\mathrm{ext}
\left[ z_\mathrm{pc}^\mathrm{ext}(V) - z_\mathrm{pcb}^\mathrm{ext}  \right]
- \frac{ \alpha^\mathrm{ext} }{  \beta_\mathrm{cb}^\mathrm{ext} }
V_\mathrm{b}^\mathrm{ext'}
\left[ z_\mathrm{pc}^\mathrm{ext}(V) - z_\mathrm{pcb}^\mathrm{ext}  \right]
\nonumber \\ && \mbox{ }
-
\frac{\alpha}{k_\mathrm{eff}} F_\mathrm{b}^\mathrm{ext'}
\left[ z_\mathrm{pc}^\mathrm{ext}(V) - z_\mathrm{pcb}^\mathrm{ext}  \right]
+ \frac{ \alpha^\mathrm{ext} }{  \beta_\mathrm{cb}^\mathrm{ext} }
V_\mathrm{b}^\mathrm{ext'}
\left[ z_\mathrm{p} - z_\mathrm{pb}^\mathrm{ext} \right]
\nonumber \\ \nonumber && \mbox{ }
+ \frac{\alpha}{k_\mathrm{eff}} F_\mathrm{b}^\mathrm{ext'}
\left[ z_\mathrm{p}- z_\mathrm{pb}^\mathrm{ext} \right]
+
\theta_\mathrm{bf}^\mathrm{ext} .
\end{eqnarray}

\subsection{Cantilever Characteristics} \label{Sec:CantChar}

For a cantilever of length $L_0$
(more precisely, the probe is attached a distance $L_0$ from the base
of the cantilever),
inclined at an angle to the horizontal of $\theta_0 < 0$,
with colloid probe of radius $R$
and diameter $L_2$,
(more precisely, $L_2 = R \sqrt{2 + 2 \cos \theta_0}$),
one can define $C_0 \equiv  \cos \theta_0$ and $S_0 \equiv  \sin \theta_0$.
The ratio of change in deflection angle to the change in vertical deflection
in the linear cantilever regime is\cite{Attard12,Attard98,Attard99}
\begin{eqnarray}
\lefteqn{
\alpha(\mu)  \equiv
\frac{\theta}{z_\mathrm{c}}
} \nonumber \\ & = &
\left\{ D(\mu) C_0
+ L_2 S_0 - \frac{R^2 S_0^2 \theta_0}{L_2} + L_2  C_0 \theta_0
\right\}^{-1} .
\end{eqnarray}
The friction coefficient is positive, $\mu > 0$.
There are three values of the ratio:
extension in contact, retraction in contact, and non-contact.
These are denoted
\begin{equation}
\alpha^\mathrm{ext} \equiv \alpha(\mu)
, \;
\alpha^\mathrm{ret} \equiv \alpha(-\mu)
,\mbox{ and }
\alpha\equiv \alpha(0).
\end{equation}
The ratio of the deflection to the deflection angle
that appears here is
\begin{eqnarray} \label{Eq:Dmu}
D(\mu) & \equiv &
\frac{x}{\theta}
\nonumber \\ & = &
\frac{2 L_0^3 [ C_0 + \mu S_0 ] + 3 L_0^2 L_2 [ S_0 - \mu C_0 ]
}{ 3 L_0^2 [ C_0 + \mu S_0 ] + 6 L_0 L_2 [ S_0 - \mu C_0 ] } .
\end{eqnarray}

\subsection{Spring Constant Determination} \label{Sec:keffb}

As discussed previously,\cite{Attard12}
the effective spring constant $k_\mathrm{eff}$
and the intrinsic spring constant $k_0$
are related by
\begin{eqnarray} \label{Eq:keff}
k_\mathrm{eff} & = &
\frac{1}{
D(0) C_0
+ L_2 S_0 + L_2  C_0 \theta_0 - {R^2 S_0^2 \theta_0}/{L_2}
 }
\nonumber \\ & & \mbox{ } \times
\frac{2 L_0  /3}{ C_0 + 2 L_2 S_0 /L_0 } k_0
\end{eqnarray}
The effective spring constant results after taking into account the
cantilever tilt and the torque on the colloid probe.
It gives directly the force acting on the probe
from the vertical deflection of the  contact point,
\begin{equation}
F_z = k_\mathrm{eff} z_\mathrm{c} .
\end{equation}
(The  vertical deflection of the  contact point
arises in the atomic force microscope force measurement
in the conversion factor for voltage,
because the calibration of this is based on the fact
that the vertical deflection of the  contact point
is equal and opposite to the movement of the piezo-drive
when the probe is in contact with the substrate.)
The intrinsic spring constant is a material property
of the cantilever and is what is measured
in, for example, the thermal calibration method.

The results given above convert the raw atomic force microscope voltage
into vertical cantilever deflection, so that one has the data pairs
$\{z_\mathrm{p},z_\mathrm{c}\}$ for both extension and retraction.
The value of the spring constant only entered this conversion
in the calculation of the base line force
$F_\mathrm{b}(z_\mathrm{p})$.
This has only a weak influence on the results.
So what one does in practice is first estimate an initial value
$ k_\mathrm{eff}^{(0)} $,
then use the following procedure to fit
the effective spring constant to the $\{z_\mathrm{p},z_\mathrm{c}\} $ data.
One can use this fitted value to replace the initial value used for the
base line force,
and then repeat the fit.
In practice, repeating the fit made negligible change.

To fit the effective spring constant
the iterative expression for the force given above
is used but for the deflection rather than the force,
\begin{equation}
z_\mathrm{c}^{(0)} =
\frac{k_\mathrm{eff}^{-1} C \dot z_\mathrm{p}}{z_0+z_\mathrm{p}} ,
\mbox{ and }
z_\mathrm{c}^{(n)} =
\frac{k_\mathrm{eff}^{-1} C \dot z_\mathrm{p} - (z_\mathrm{c}^{(n-1)})^2
}{
z_0+z_\mathrm{p} + z_\mathrm{c}^{(n-1)} }.
\end{equation}
In practice two iterates were used,
which means three columns of data
each for extension and retraction,
with each row corresponding to a given measured $z_\mathrm{p} $.
Note that this is the stick drainage force formula;
the slip length does not come into it.

It should be mentioned that this ignores the rate
of change of the cantilever deflection, $\dot z_\mathrm{c}$.
This is valid at large separations where
$|\dot z_\mathrm{c}| \ll |\dot z_\mathrm{p}|$.

The calculated deflection was converted to voltage by using
\begin{equation}
V^\mathrm{ext}(z_\mathrm{p}) =
\beta_\mathrm{cb}^\mathrm{ext} \left[
k_\mathrm{eff}^{-1}  F_\mathrm{b}^\mathrm{ext}(z_\mathrm{p})
- z_\mathrm{c} \right]
+ V_\mathrm{b}^\mathrm{ext}(z_\mathrm{p}) ,
\end{equation}
and similarly for retraction.
The linear base line force that appears here
is the one with the original estimate of the effective spring constant.
The use of the base line calibration factor $\beta_\mathrm{cb}$
means that this is restricted to the linear regime,
which means that the deflections cannot be too large,
which again restricts the formula to large separations.
The square of the difference between this voltage
and the measured raw voltage was summed
over the specified ranges for extension and retraction
and minimised to obtain $k_\mathrm{eff}$.

The iteration procedure fails for large deflections.
Depending upon the system, large can mean 1--10$\,$nm.
It is clear when it fails, because the iterations do not converge.
Care has to be taken to exclude failed $z_\mathrm{c} $ from the error estimate.
The way this was done is to specify a range
$[z_\mathrm{p}^\mathrm{small},z_\mathrm{p}^\mathrm{large}]$
over which the errors were summed.
The upper limit is close to the start where the piezo-drive
is moving at uniform velocity; it is typically
the same at the base line upper limit.
The lower limit is greater than the first position when the iteration
procedure fails.
Typically a range of 3--5$\,\mu$m was used for the estimate of the error
in the fit.
One can always tell if the range is appropriate and if the fit is good
from a plot of the measured and the calculated deflections.

%
\section{Variable Drag Derivation and Algorithm}
%

\subsection{Drag Length}

At large separations the drag force on the cantilever is constant
and can be described by an effective drag length,
\begin{equation}
F_\mathrm{drag}^\mathrm{ext}
= - 6 \pi \eta \dot z_\mathrm{p}^\mathrm{ext} L_\mathrm{drag}
= -F_\mathrm{drag}^\mathrm{ret} .
\end{equation}
The effective drag length is typically a fraction of the length
of the cantilever,
but this obviously varies between cantilevers.

The constant drag force is removed from the measured experimental data
by the treatment of the base line discussed above.
Hence in most cases one does not need to know directly the drag force
or the effective drag length.
However in some cases,
particularly for soft cantilevers,
highly viscous liquids,
and high driving velocities,
the variation in the cantilever drag force with cantilever deflection
cannot be neglected.
In this case the effective drag length is needed
in order to calculate the variable drag force,
as is discussed shortly.

The difference in the base line voltage on extend and retract
is due in part to the change in sign of the constant drag force,
in part due to the change in sign of the drainage force
in the base line region,
and in part due to thermal drift.
The latter two have to be subtracted in order to obtain
the drag force alone.
The change in voltage due only to thermal drift is
\begin{eqnarray}
\Delta V_\mathrm{therm}
& = &
\frac{1}{2}
\left\{
\left[
V_\mathrm{b}^\mathrm{ext'} - V_\mathrm{b}^\mathrm{ret'} \right]
\right.
\nonumber \\ && \left. \mbox{ }
-
\left[
F_\mathrm{b}^\mathrm{ext'} - F_\mathrm{b}^\mathrm{ret'} \right]
k_\mathrm{eff}^{-1} \lambda(V_\mathrm{b})
\right\}
\nonumber \\ && \mbox{ } \times
\left[ 2 z_\mathrm{p,turn}
- z_\mathrm{pb}^\mathrm{ext} - z_\mathrm{pb}^\mathrm{ret}  \right] .
\end{eqnarray}
where $z_\mathrm{p,turn}$ is the furthest position in contact,
where the piezo-drive turns from extension to retraction,
and $V_\mathrm{b} \equiv
[V_\mathrm{b}^\mathrm{ext} + V_\mathrm{b}^\mathrm{ret}]/2$.
Clearly the contribution to the change in voltage from the drainage force
is being subtracted here.

This expression assumes that the piezo-drive velocity is constant
on each branch. Since in fact it slows just before the turn point,
and it accelerates just after,
it would be better to write this in terms of the change in time
(since the thermal drift is constant in time).
In practice this was found to make negligible difference in the drag force.

With the above expression for the change in voltage
due to drift, the drag force is
\begin{eqnarray}
F_\mathrm{drag}^\mathrm{ext}
& = &
\left[
V_\mathrm{b}^\mathrm{ext} - V_\mathrm{b}^\mathrm{ret}
- \Delta V_\mathrm{therm} \right]
\frac{k_\mathrm{eff}}{ 2\lambda(V_\mathrm{b})}
\nonumber \\ &&  \mbox{ }
-
\frac{1}{2} \left[
F_\mathrm{b}^\mathrm{ext} - F_\mathrm{b}^\mathrm{ret} \right] ,
\end{eqnarray}
where again the drainage force is subtracted.
In practice the term $ \Delta V_\mathrm{therm} $ is often negligible.
From this the effective drag length can be obtained.
The standard deviation in the effective drag length
over all velocities is about 3\%.

There is an alternative way of obtaining the effective drag length
that also works well and serves as a useful check
or replacement for the above procedure.
At the beginning of the extension run,
the piezo-drive decelerates from
$\dot z_\mathrm{p}^\mathrm{ext} = 0$ to
$\dot z_\mathrm{p}^\mathrm{ext} = - |\dot z_\mathrm{p}|$
over typically a distance of several hundred nanometers.
Similarly, at the end of the retraction run it
accelerates from  $\dot z_\mathrm{p}^\mathrm{ret} = |\dot z_\mathrm{p}|$ to
$\dot z_\mathrm{p}^\mathrm{ret} = 0$.
Depending on the model of the atomic force microscope
and the settings,
this data is recorded
and appears as a distinct region at the terminus of the extended
flat plateaux that represent the drag and drainage forces
for constant velocity.
The  change in voltage (equivalently force) from the beginning
on extension, or end on retraction
to the plateaux values gives the drag force
(after the asymptotic value of the drainage force has been subtracted).
This second method is almost identical to the first method
neglecting  $\Delta V_\mathrm{therm}$.

\subsection{Drainage and van der Waals Force} \label{Sec:FvdW-drain}

The drainage force is
\begin{equation}
F_\mathrm{drain}(h,\dot h)
=
\frac{-6\pi\eta R^2 \dot h}{ h} v(h).
\end{equation}
For the case of stick, $v(h) = 1$.
In this case the force is attributed to Taylor.
For slip and symmetric surfaces each with slip length $b$ on each surface,
the Vinogradova factor is\cite{Vinogradova95}
\begin{equation}
v(h) =
\frac{h}{3b} \left\{ \left[ 1 + \frac{h}{6b}  \right]
\ln\left[ 1 + \frac{6b}{h}  \right] - 1 \right\} .
\end{equation}
For the asymmetric case,
with slip lengths $b_1$ and $b_2$,
one defines the relative asymmetry as
$k \equiv [b_2-b_1]/b_1$ and also $A \equiv b_1+b_2$,
\begin{equation}
C_\pm \equiv 2 b_1 \left[ 2 + k \pm \sqrt{ 1+k+k^2 } \right] .
\end{equation}
With these the factor is\cite{Vinogradova95}
\begin{eqnarray}
v(h) & = &
\frac{-2Ah}{C_+C_-}
-
\frac{2h}{C_--C_+}
\nonumber \\ && \mbox{ } \times
\left\{
\frac{(C_++h)(C_+-A)}{C_+^2} \ln\left[ 1 + \frac{C_+}{h}  \right]
\right. \nonumber \\ && \left. \mbox{ }
-
\frac{(C_-+h)(C_--A)}{C_-^2} \ln\left[ 1 + \frac{C_-}{h}  \right]
\right\} .
\end{eqnarray}
In both case slip is negligible at large separations,
$v(h) \rightarrow 0$ as $h \rightarrow \infty$.

The van der Waals force is taken to be\cite{Attard92}
\begin{equation}
F_\mathrm{vdw}(h) = \frac{-AR}{6h^2}
\left[1 - \frac{1}{4} \left( \frac{z^\dag}{h} \right)^6  \right] .
\end{equation}
Here $A$ is the Hamaker constant.
This includes a short ranged repulsive force,
which is derived from a Lennard-Jones 6-12 potential.
The location of the minimum in the potential
is here taken to be $z^\dag = 0.53\,$nm.
The van der Waals force in general has no effect on the drainage force
or the slip length on approach (extension).
It does however effect the adhesion, as is shown in the results
in the text.
For convenience the sum of the point forces acting on the probe
will be denoted
$F_z \equiv F_\mathrm{vdw} + F_\mathrm{drain}$.

For use below, the derivative of the van der Waals force is
\begin{equation}
\frac{\partial F_\mathrm{vdW}}{\partial h }
=\frac{AR}{3h^3}
\left[1
- \left( \frac{z^\dag}{h} \right)^6  \right] ,
\end{equation}
and those for the drainage force are
\begin{equation}
\frac{\partial F_\mathrm{drain}}{\partial h }
=
6\pi\eta R^2 \dot h
\left[ \frac{v(h)}{ h^2} - \frac{v'(h)}{ h} \right] ,
\end{equation}
where the prime denotes the derivative with respect to separation,
and
\begin{equation}
\frac{\partial F_\mathrm{drain}}{\partial \dot h }
=
\frac{-6\pi\eta R^2}{h} v(h) .
\end{equation}

\subsection{Cantilever Analysis} \label{Sec:Analysis}

\begin{figure}[t!]
\centerline{
\resizebox{8.5cm}{!}{ \includegraphics*{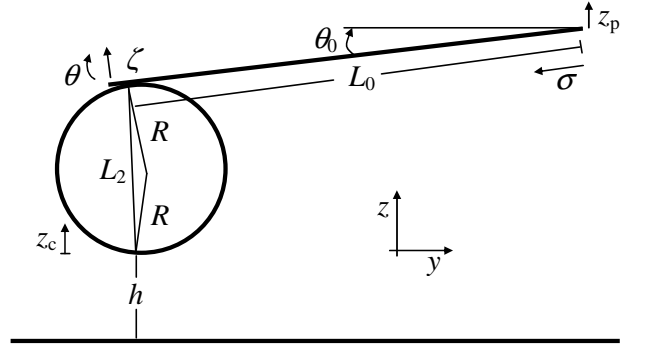} } }
\caption{\label{Fig:AFMgeom}
Cantilever geometry in the atomic force microscope
(not to scale).
}
\end{figure}

This analysis follows closely earlier analysis
of the bending of an atomic force microscope cantilever
by Attard and co-workers.\cite{Attard98,Attard99,Zhu11a,Attard12}
In turn, that analysis is based on
the classical equations of continuum elasticity.\cite{Southwell36}

Consider a rectangular cantilever
of length $L_0$ and width $w \ll L_0$,
tilted at an angle of $\theta_0 < 0$ to the horizontal
with long axis in the $yz$-plane.
Define $C_0 \equiv \cos \theta_0$ and $S_0 \equiv \sin \theta_0$.
The tilt angle is small, but no linearisation will be performed.
Angles are measured in a clockwise direction,
and the $y$-coordinate is measured left to right.
There is a spherical colloid probe of radius $R$ attached
to the end of the cantilever.
There is a planar substrate in the $xy$-plane located
beneath the cantilever
such that zero separation for the undeflected cantilever
occurs when the base of the cantilever is at $z_\mathrm{p}=0$.

Let $\sigma \in [0,L_0]$ denote the running coordinate
along the cantilever, with $\sigma = 0$ being the base
and $\sigma = L_0$ being the end.
Note that this runs right to left,
the opposite sense to the $y$-coordinate.
Let $\zeta(\sigma)$ denote the deflection orthogonal
to the original axis,
with $\zeta_L \equiv \zeta(L)$ being the deflection of the end.
The deflection angle is
$\theta(\sigma) = \mathrm{d} \zeta(\sigma) /\mathrm{d} \sigma$,
or, equivalently,
\begin{equation}
\zeta(\sigma)  = \int_0^\sigma \mathrm{d} \sigma' \, \theta(\sigma') .
\end{equation}
The boundary condition is
$ \zeta(0) = \theta(0) = 0$.
The deflection angle of the end, $\theta_L \equiv \theta(L)$,
is what is measured in the atomic force microscope.
The deflection will be assumed small
and everything will be linearised with respect to it.

In the laboratory frame,
the undeflected cantilever
is described by
\begin{equation}
y_0(\sigma) = -C_0 \sigma
, \mbox{ and }
z_0(\sigma) = z_\mathrm{p} + S_0 \sigma.
\end{equation}
Here $z_\mathrm{p}$ is the piezo-drive position,
which is the location of the base of the cantilever.

The deflected cantilever has laboratory coordinates
\begin{eqnarray}
y(\sigma) & = &
-\int_0^\sigma \mathrm{d} \sigma' \, \cos(\theta_0 + \theta(\sigma') )
\nonumber \\ & = &
-C_0 \sigma
+ S_0 \int_0^\sigma \mathrm{d} \sigma' \, \theta(\sigma')
+ {\cal O}(\theta^2)
\nonumber \\ & = &
y_0(\sigma) + S_0 \zeta(\sigma),
\end{eqnarray}
and
\begin{eqnarray}
z(\sigma) & = &
z_\mathrm{p}
+ \int_0^\sigma \mathrm{d} \sigma' \, \sin(\theta_0 + \theta(\sigma') )
\nonumber \\ & = &
z_\mathrm{p} + S_0 \sigma
+ C_0 \int_0^\sigma \mathrm{d} \sigma' \, \theta(\sigma')
+ {\cal O}(\theta^2)
\nonumber \\ & = &
z_0(\sigma) + C_0 \zeta(\sigma).
\end{eqnarray}

With total angle $\theta_\mathrm{tot} = \theta_0 + \theta$,
a colloid probe of radius $R$ has lever arm
\begin{eqnarray}
L_2(\theta_\mathrm{tot})
& = &
R\sqrt{ 2 + 2 \cos \theta_\mathrm{tot} }
\nonumber \\ & = &
R \sqrt{ 2 + 2 C_0 - 2 S_0 \theta + {\cal O}(\theta^2) }
\nonumber \\ & = &
L_2
\left[ 1 - \frac{ S_0 \theta}{2[1+C_0]} \right] + {\cal O}(\theta^2)
\nonumber \\ & = &
L_2 - \frac{ R^2 S_0 \theta}{L_2}
+ {\cal O}(\theta^2) ,
\end{eqnarray}
where $L_2 \equiv L_2(\theta_0) =  R \sqrt{ 2 + 2 C_0 }$.
The change in the length of the lever arm with deflection angle is small.

The change in the point of closest approach
of the colloid probe (contact position) is
horizontally
\begin{eqnarray}
y_\mathrm{c} & = &
S_0 \zeta_L
-  L_2(\theta_\mathrm{tot}) \cos(\theta_\mathrm{tot}) \theta_\mathrm{tot}
+ L_2(\theta_0) \cos(\theta_0) \theta_0
\nonumber \\ & = &
S_0 \zeta_L
-  \left[ L_2 C_0
-  L_2 S_0 \theta_0
- C_0 \theta_0 \frac{ R^2 S_0 }{L_2} \right] \theta_L ,
\end{eqnarray}
and vertically
\begin{eqnarray} \label{Eq:zc}
z_\mathrm{c} & = &
C_0 \zeta_L
+ \sin(\theta_\mathrm{tot}) L_2(\theta_\mathrm{tot}) \theta_\mathrm{tot}
- S_0 L_2 \theta_0
\nonumber \\ & = &
C_0 \zeta_L
+ \left[ S_0 + C_0 \theta_0
- S_0  \theta_0 \frac{ R^2 S_0 }{L_2^2} \right] L_2 \theta_L
\nonumber \\ & \equiv &
C_0 \zeta_L + S_0^* L_2 \theta_L .
\end{eqnarray}
These are given to linear order in the deflection
and represent the change from the position of the undeflected cantilever.
Note that to leading order
the change in horizontal position
has the opposite sign to the deflection and the deflection angle.
This determines the sign of the friction force below.

The separation is
\begin{equation}
h = z_\mathrm{p} + z_\mathrm{c} .
\end{equation}
As mentioned above,
the base of the cantilever has been positioned relative to the substrate
so that the undeflected cantilever
is in contact when  $ z_\mathrm{p} = 0$.

Let $F_z$ and $F_y$ be the components of the force acting
acting on the colloid probe at the point of closest approach.
The van der Waals and drainage force on the colloid probe
are represented by $F_z$,
and the friction force, when present, is represented by $F_y$.
Typically, $F_y = \mu F_z$
in contact on extension, where $\mu$ is the friction coefficient
(see the above comment regarding the sign of the change in horizontal position).
These forces create a force on the end of the cantilever,
\begin{equation}
F_L = F_z C_0 + F_y S_0 ,
\end{equation}
which is orthogonal to the axis of the cantilever,
and a turning moment (torque),
\begin{equation}
M_L = L_2 F_z S_0 - L_2 F_y C_0 ,
\end{equation}
which is positive in the clockwise direction.
In these $\theta_\mathrm{tot} = \theta_0 + \theta$
has been replaced by $\theta_0$ because only terms linear
in the deflection are retained.

Now add a force per unit length $f(\sigma)$
in the cantilever frame
(i.e.\ acting orthogonal to the undeflected cantilever).
This is the drag (and possibly drainage) force on the cantilever,
which will be given explicitly in the following section.
In the cantilever frame,
the turning moment about $\sigma$
due to the end force, end moment, and force per unit length is
\begin{equation}
M(\sigma) = [L_0-\sigma] F_L + M_L
+ \int_\sigma^{L_0} \mathrm{d} \sigma' \, f(\sigma') [ \sigma'-\sigma ] .
\end{equation}
From this the deflection of the cantilever is\cite{Southwell36}
\begin{eqnarray} \label{Eq:shape}
\zeta(\sigma) & = &
B^{-1}  \int_0^\sigma \mathrm{d} \sigma'' \,
M(\sigma'') [ \sigma-\sigma'' ]
\nonumber \\  & = &
\frac{F_L}{B}  \left[ \frac{L_0 \sigma^2 }{2} - \frac{\sigma^3}{6}  \right]
+  \frac{\sigma^2 M_L}{2B}
\nonumber \\ & & \mbox{ }
+ \frac{1}{B}  \int_0^\sigma  \mathrm{d} \sigma' \, f(\sigma')
\left[ \frac{ \sigma \sigma'\!\,^2}{2}
- \frac{\sigma'\!\,^3}{6}  \right]
\nonumber \\  & & \mbox{ }
+ \frac{1}{B}  \int_\sigma^{L_0} \mathrm{d} \sigma' \, f(\sigma')
\left[ \frac{ \sigma^2 \sigma'}{2}
- \frac{\sigma^3}{6}  \right] .
\end{eqnarray}
Here $B = EI$ is the beam elasticity parameter,
which is related to the intrinsic spring constant of the cantilever by
$k_0 = 3B/L_0^3$.

For the case of a uniform distributed force, $f(\sigma) = f$,
\begin{eqnarray} \label{Eq:zeta-f-uni}
\zeta_L & = &
\frac{F_L}{B}  \frac{L_0^3  }{3}
+  \frac{L_0^2 M_L}{2B}
+ \frac{f}{B} \left[ \frac{L_0^4}{6} - \frac{L_0^4}{24} \right]
\nonumber \\ & = &
\frac{F_L}{k_0} +  \frac{3 M_L}{2 L_0 k_0 }
+ \frac{3 L_0 f }{8 k_0}  .
\end{eqnarray}
For the case that there is no drag or drainage force on the cantilever,
$f(\sigma) = 0$,
the end deflection is
\begin{equation}
\zeta_L
= \frac{F_L}{k_0} +  \frac{3 M_L}{2 L_0 k_0 }.
\end{equation}
If in addition there is no lever arm, $L_2 = M_L=0$,
nor friction force, $F_y=0$,
so that the end force is $F_L = F_z C_0$,
then the end deflection reduces to
\begin{equation} \label{Eq:zetaL}
\zeta_L
= \frac{F_z C_0}{k_0}
,\mbox{ or }
z_\mathrm{c} = \frac{F_z C_0^2 }{k_0} .
\end{equation}
Here $\zeta_L C_0 $ is the vertical deflection of the end of the cantilever
in the laboratory frame.
In this case with no lever arm, $L_2 = 0$,
the vertical deflection of the end is the same as the change in
vertical position of the contact point, $z_\mathrm{c}=\zeta_L C_0$.

This result does not agree with the result given
by Vinogradova and Yakubov\cite{Vinogradova03} for the same case
(their Eq.~(6), in the present notation, is
$ z_\mathrm{c} = {F_z} [ 3 C_0 - 1 ] /{2k_0}$).
In the opinion of the present author,
the error in Ref.~\onlinecite{Vinogradova03}
arises in the very first equation
where the authors mix up the internal coordinates of the cantilever
with the laboratory coordinates.
Consequently, the present author believes
that all of the results in Ref.~\onlinecite{Vinogradova03} are suspect.

From the general expression for the deflection
of the cantilever, the angular deflection of the end,
$\theta_L = \left. \mathrm{d} \zeta(\sigma) / \mathrm{d}\sigma
\right|_{\sigma = L_0}$, is
\begin{equation}
\theta_L =
\frac{L_0^2 F_L}{2B} + \frac{L_0 M_L}{B}
+ \frac{1}{B}  \int_0^{L_0}  \mathrm{d} \sigma' \, f(\sigma')
\frac{\sigma'\!\,^2}{2}  .
\end{equation}
This is important because it is what is actually measured
in the atomic force microscope.
For the case of a uniform force per unit length, $f(\sigma) = f$,
this reduces to
\begin{equation} \label{Eq:QL-unif}
\theta_L =
\frac{L_0^2 F_L}{2B} + \frac{L_0 M_L}{B}
+ \frac{L_0^3 f}{6B}  .
\end{equation}

In the case of only probe forces,
(i.e.\ no distributed drag or drainage forces on the cantilever),
a number of useful constants can be given.
First define for convenience
\begin{equation}
C_+ \equiv C_0 + \tilde \mu S_0
\mbox{ and }
S_- \equiv S_0 - \tilde \mu C_0,
\end{equation}
which differ from $C_0$ and $S_0$ only in contact
since the effective friction coefficient is defined as
\begin{equation}
\tilde \mu
=
\left\{
\begin{array}{ll}
\mu , & \mbox{contact, extension}, \\
-\mu, & \mbox{contact, retraction}, \\
0, & \mbox{non-contact} .
\end{array} \right.
\end{equation}
With these,
the ratio of the end deflection to the surface force is
\begin{equation}
\frac{\zeta_L}{F_z} =
\frac{C_+}{k_0}
+ \frac{3L_2S_-}{2L_0k_0},
\end{equation}
and the ratio of the angle deflection to the surface force is
\begin{equation}
\frac{\theta_L}{F_z} =
\frac{3C_+}{2L_0k_0}
+ \frac{3L_2S_-}{L_0^2k_0} .
\end{equation}
Using these,
the effective spring constant,
which  is defined as the ratio of the vertical force
to the vertical contact point deflection,
is
\begin{eqnarray} \label{Eq:keff}
\lefteqn{ k_\mathrm{eff}(\tilde \mu)
\equiv  \frac{F_z}{z_\mathrm{c}}
} &&
 \\ & = &
\frac{F_z}{C_0 \zeta_L + S_0^* L_2 \theta_L }
\nonumber \\ & = & \nonumber
\frac{k_0}{C_0 \left[ C_+ + \frac{3L_2S_-}{2L_0} \right]
+
S_0^* L_2 \left[ \frac{3C_+}{2L_0}
+ \frac{3L_2S_-}{L_0^2} \right]} .
\end{eqnarray}
When one speaks of `the' effective spring constant,
one means out of contact, $k_\mathrm{eff} \equiv k_\mathrm{eff}(0)$,
(which means $C_+ = C_0$, $S_-=S_0$).
One can see that for the case $L_2=0$,
which means that $z_\mathrm{c}=C_0 \zeta_L$,
this is the same as Eq.~(\ref{Eq:zetaL}).
These expression agree with those given in \S\S \ref{Sec:CantChar}
and \ref{Sec:keffb}.

These three equations are sufficient to obtain
everything that is needed
from an atomic force measurement of surface forces
that has been analysed to give the angular deflection.
That is, they give $F_z(\theta_L)$,
and hence $z_\mathrm{c}(F_z)$,
and hence $h = z_\mathrm{p} - z_\mathrm{c}(F_z) + z_0$.
This assumes that the the angular deflection
and the zero of separation have been obtained
as described in \S\ref{Sec:AppB}.
It also assumes that any variation in distributed forces
on the cantilever can be neglected,
which means that these are apparent rather than real quantities.

\subsection{Drag Force on Cantilever}

In the atomic force microscope
one measures a drag force at large separations
\begin{equation}
F_\mathrm{drag} = -6 \pi \eta L_\mathrm{drag} \dot z_\mathrm{p} .
\end{equation}
This defines the drag parameter $ L_\mathrm{drag}$.
The drag length is an effective quantity
that accounts for the cantilever shape,
the fact that the drive velocity is not orthogonal
to the cantilever axis,
(if one wanted to, one could instead define
$F_\mathrm{drag} = -6 \pi \eta \tilde L_\mathrm{drag} C_0 \dot z_\mathrm{p} $,
with $\tilde L_\mathrm{drag}  = L_\mathrm{drag} /C_0$),
and, most importantly,
the fact that the distributed drag force
is interpreted as if it were a point force on the contact point.

The drag force per unit length may be taken to be
\begin{equation} \label{Eq:fsigma}
f(\sigma)
=
- c_\mathrm{drag} [ \dot z_\mathrm{p} C_0 + \dot \zeta(\sigma) ] .
\end{equation}
This simply says that the local drag force is proportional
to the local velocity of the cantilever through the liquid.
This drag force is orthogonal to the axis of the undeflected cantilever,
which was how the distributed force was defined.

At large separations $\dot \zeta(\sigma) \approx 0$
and this force is uniform.
In this case,
Eq.~(\ref{Eq:QL-unif}) gives the deflection angle as
\begin{equation}
\theta_L =
\frac{-c_\mathrm{drag} \dot z_\mathrm{p} C_0  }{2 k_0}  .
\end{equation}

The end force and moment
due to a point force, $F_z = F_\mathrm{drag}$,
and no friction, $F_y=0$, is
$F_L =  C_0 F_\mathrm{drag} $ and $M_L = L_2 S_0  F_\mathrm{drag}$,
which gives a deflection angle
\begin{equation}
\theta_L =
\left[  \frac{3  C_0 }{2L_0 k_0} + \frac{3 L_2 S_0 }{L_0^2 k_0} \right]
F_\mathrm{drag} .
\end{equation}
Equating these two expressions for the deflection angle
gives
\begin{eqnarray}
c_\mathrm{drag}  & = &
\frac{2 k_0}{\dot z_\mathrm{p} C_0}
\left[  \frac{3  C_0 }{2 L_0 k_0} + \frac{3 L_2 S_0 }{L_0^2 k_0} \right]
6 \pi \eta L_\mathrm{drag} \dot z_\mathrm{p}
\nonumber \\ & = &
\frac{12 \pi \eta L_\mathrm{drag}}{ C_0}
\left[  \frac{3  C_0 }{2 L_0} + \frac{3 L_2 S_0 }{L_0^2} \right] .
\end{eqnarray}
The leading term is identical to that used previously.\cite{Zhu11a}
(Previously the subordinate term
due to tilt and torque were not accounted for.)


\subsection{Comparison of Theory and Measurement}

Next an algorithm  will be given
for computing the deflection of the cantilever
due to the drainage and van der Waals forces on the  colloid probe
and the drag force on the cantilever.
This algorithm gives the actual vertical deflection $z_\mathrm{c}$,
the actual separation $h$, and the actual surface force $F_z$,
as well as the actual angular deflection $\theta_L$.
The atomic force microscope measures  $\theta_L$,
and assumes that this angular deflection is due solely
to a point force $F_z$, and uses the point force equations
to convert the angular deflection to an apparent vertical deflection,
an apparent separation, and an apparent point force.
This procedure would be accurate if the distributed drag force
on the cantilever were constant,
(because then the distributed force contribution to the deflection
is just a constant that is removed in the base line treatment of the data).
However to the extent that the drag force is varying with
cantilever deflection,
the apparent quantities are not equal to the actual quantities.
Therefore,
in comparing theory with measurement
one has to convert an exact angular deflection given by theory
to  an apparent vertical deflection,
an apparent separation, and an apparent point force,
using the same conversion factors that one uses
to analyze the atomic force microscope raw data.
(Out of interest, one can always compare
the actual and the apparent calculated quantities
to see how big an effect variable drag has.)
In the text above, it is the apparent theoretical and experimental
quantities that are plotted.

%
%

\subsection{Numerical Computation for Variable Drag}

The shape of the cantilever $\zeta(\sigma)$
under the influence of the point and distributed forces
was given above as Eq.~(\ref{Eq:shape}).
The shape is a function $F_z$
(via the friction force $F_y$, if present,
the end force $F_L$, and the end moment $M_L$),
which gives it a direct dependence  on $h$ and $\dot h$,
and a functional of $\dot \zeta(\sigma)$ (via $f(\sigma)$).
This function will be denoted $S(\sigma;F_z,[\zeta])$,
or more simply $S(\sigma)$.
(This is necessary
because in the numerical solution of the equations of motion,
one has to distinguish and to equate
the shape that evolves from the previous shape, $\zeta(t+\Delta_t)$,
and the shape that satisfies the elasticity equation,
$S(\sigma(t+\Delta_t))$.)
The deflection of the end for a given shape is
$\zeta_L \equiv \zeta(L_0) = S(L_0)$.
Similarly the deflection angle of the end is
$\theta_L = \left. \mathrm{d} S(\sigma)/\mathrm{d} \sigma
\right|_{\sigma=L_0}$.
The separation is $h = z_\mathrm{p} + z_\mathrm{c}$,
and the rate of change of separation is
$\dot h = \dot z_\mathrm{p} + \dot z_\mathrm{c}$,
where $z_\mathrm{c}$ is the function of $\zeta_L$
and $\theta_L$ given above, Eq.~(\ref{Eq:zc}).
For convenience the constant of separation has been chosen as zero.

The problem to be solved is: given the drive trajectory $z_\mathrm{p}(t)$,
at each instant calculate the shape $S(\sigma;t)$
and hence $\zeta_L(t)$, $\theta_L(t)$, $h(t)$ etc.
The axial coordinate can be discretised,
$\sigma_i = i\Delta_\sigma$ and the shape can be written simply as a vector
${\bm \zeta}$, with elements $\zeta_i = \zeta(\sigma_i)$.
Also $i \in [0,L]$,
$\sigma_L = L_0$ and $\zeta_L = \zeta(y_L) = \zeta(L_0)$,
consistent with the notation introduced above.

Suppose the shape ${\bm \zeta}$ and its rate of change
$\dot{\bm \zeta}$ is known at time $t$.
It is required to obtain these at time $t' = t+\Delta_t$.
One has three equations to solve:
\begin{eqnarray}
{\bm \zeta}' & = & {\bm \zeta}
+ \Delta_t \dot{\bm \zeta}
+ \frac{1}{2} \Delta_t^2 \ddot{\bm \zeta},
\nonumber \\
\dot{\bm \zeta}' & = & \dot{\bm \zeta}
+ \Delta_t  \ddot{\bm \zeta},
\nonumber \\
{\bm \zeta}' & = & {\bf S}(F_z',\dot{\bm \zeta}').
\end{eqnarray}
Clearly then one has a single unknown vector,
namely the acceleration $\ddot{\bm \zeta}$,
and one can insert the first two equations into the third,
which can then be written as a function of the acceleration,
${\bm \zeta}'(\ddot{\bm \zeta})
= {\bf S}(\ddot{\bm \zeta})$.
(Both sides of this depend upon the known ${\bm \zeta}$ and $\dot {\bm \zeta}$
at the preceding time step.)
Equivalently then one has to find the zero of the vector function
\begin{equation}
{\bf G}(\ddot{\bm \zeta})
\equiv
{\bm \zeta}'(\ddot{\bm \zeta})
- {\bf  S}(\ddot{\bm \zeta}) .
\end{equation}
In order to develop a stable iteration procedure to solve this,
one converts it to a fixed point problem
and uses essentially Newton's method.
That is, define
\begin{equation}
{\bf H}(\ddot{\bm \zeta})
\equiv
\ddot{\bm \zeta}
- \left(
\frac{\partial {\bf G} }{\partial \ddot{\bm \zeta}} \right)^{-1}
{\bf G}(\ddot{\bm \zeta}) ,
\end{equation}
where the Jacobean matrix has elements
$\left\{ {\partial {\bf G} }/{\partial \ddot{\bm \zeta}}\right\}_{ij}
= {\partial G_i }/{\partial \ddot { \zeta}_j}$.
Clearly the correct acceleration $\overline {\ddot{\bm \zeta}}$,
which satisfies ${\bf G}(\overline {\ddot{\bm \zeta}})
= {\bf 0}$
is a fixed point of this, $ {\bf H}(\overline {\ddot{\bm \zeta}})
= \overline {\ddot{\bm \zeta}}$.
Hence one can set up an iteration procedure for the acceleration,
\begin{equation}
\ddot{\bm \zeta}^{(n+1)} =
{\bf H}(\ddot{\bm \zeta}^{(n)} ).
\end{equation}
The procedure has been designed to be quadratically convergent
by making the Jacobean of ${\bf H}$ vanish at the fixed point.
To avoid matrix inversion,
this is more conveniently written
\begin{equation}
\frac{\partial {\bf G}^{(n)} }{\partial \ddot{\bm \zeta}} \;
\ddot{\bm \zeta}^{(n+1)} =
\frac{\partial {\bf G}^{(n)} }{\partial \ddot{\bm \zeta}}  \;
\ddot{\bm \zeta}^{(n)}
-
{\bf G}^{(n)} .
\end{equation}
Care has to be taken with the matrix multiplications,
since these should be weighted according to the trapezoidal quadrature rule.
Gaussean elimination can be used to obtain from this
the next iterate for the acceleration.
For the first guess,
the converged acceleration from the preceding time step can be used,
$\ddot{\bm \zeta}^{(0)}(t') = \ddot{\bm \zeta}(t)$.

The elements of the Jacobean matrix are explicitly
\begin{equation}
\frac{\partial G_i }{\partial \ddot { \zeta}_j}
=
\frac{\Delta_t^2}{2} \delta_{ij}
-
\frac{\partial S_i' }{\partial F_z'}
\frac{\partial  F_z' }{\partial \ddot { \zeta}_j  }
-
\Delta_t  \frac{\partial S_i }{\partial \dot \zeta_j} .
\end{equation}
with $F_z' = F_z(t+\Delta_t)$.
A Kronecker-$\delta$ appears here.
One has
\begin{eqnarray}
\frac{\partial S_i }{\partial \dot \zeta_j}
& = &
\frac{-\Delta_\sigma c_\mathrm{drag} /B}{1+\delta_{j0}+\delta_{jL}}
\times
\left\{
\begin{array}{ll} \displaystyle
\frac{ \sigma_i \sigma_j^2}{2} - \frac{ \sigma_j^3}{6}
& j \le i , \\ \displaystyle
 \rule{0cm}{.6cm} 
 \frac{ \sigma_i^2 \sigma_j}{2} - \frac{ \sigma_i^3}{6} ,
& j \ge i ,
\end{array} \right.
\end{eqnarray}
and
\begin{equation}
\frac{\partial S_i }{\partial F_z}
=
B^{-1} \left[ \frac{1}{2} L_0 \sigma_i^2 - \frac{1}{6} \sigma_i^3 \right]
\frac{\partial F_L }{\partial F_z }
+
\frac{y_i^2}{2B} \frac{\partial M_L }{\partial F_z } .
\end{equation}
In turn one has
\begin{equation}
\frac{\partial F_L }{\partial F_z }
=
C_0 + \tilde \mu S_0
, \mbox{ and }
\frac{\partial M_L }{\partial F_z }
=
L_2 S_0 - \tilde \mu L_2 C_0 ,
\end{equation}
where the effective friction constant was given above.
Finally,
\begin{eqnarray}
\frac{\partial  F_z' }{\partial \ddot { \zeta}_j  }
& = &
\frac{\partial  F_z }{\partial h }
\frac{\partial  z _\mathrm{c}'}{\partial \ddot { \zeta}_j  }
+
\frac{\partial  F_z }{\partial \dot h }
\frac{\partial  \dot z _\mathrm{c}'}{\partial \ddot { \zeta}_j  }
\nonumber \\  & = &
\left[
\frac{\Delta_t^2}{2}
\frac{\partial  F_z }{\partial h }
+
\Delta_t
\frac{\partial  F_z }{\partial \dot h }
\right]
\left[
C_0 \frac{\partial  \zeta_L }{\partial \zeta_j  }
+ S_0^* L_2 \frac{\partial  \theta_L }{\partial \zeta_j  }
\right]
\nonumber \\  & = &
\left[
\frac{\Delta_t^2}{2}
\frac{\partial  F_z }{\partial h }
+
\Delta_t
\frac{\partial  F_z }{\partial \dot h }
\right]
\nonumber \\  &  &  \mbox{ } \times
\left[ C_0 \delta_{Lj}
+
\frac{S_0^* L_2}{\Delta_\sigma} \left\{ \delta_{Lj} - \delta_{L-1,j} \right\}
\right] .
\end{eqnarray}
The final term follows because
the deflection angle of the end is
$\theta_L \equiv [\zeta_L - \zeta_{L-1} ]/\Delta_\sigma $.
The force derivatives were given in \S\ref{Sec:FvdW-drain}.

Typically $L=100$ points were used to discretise the axial integrals.
This many points markedly slows the Gaussean elimination.
Therefore an orthonomal basis for a sub-space was constructed,
$\underline v_n$, with $n=1,2, \ldots$,
with elements $v_{ni}$ a polynomial in $y_i$ of degree $n$.
The polynomials are chosen orthogonal on the interval $[0,L_0]$
with trapezoidal weighting.
The orthonomal basis was constructed
numerically. Typically four vectors were used for the basis.
The various matrices and vectors were projected onto the sub-space
and the iteration for the acceleration carried out.
Typically 5 iterations were used at each time step.
Tests showed the results did not change upon increasing
the number of iterations, basis vectors, or grid points.
The procedure was found to be quite stable.
This includes in contact, the turn around point,
and the jump from contact.

\subsection{Simplified Algorithm for Constant Drag} \label{Sec:AlgoDrain}

The algorithm simplifies considerably
if one can neglect the variation in cantilever drag with deflection.
In this case one only needs the vertical deflection $z_\mathrm{c}$,
the velocity, $\dot z_\mathrm{c}$,
and the acceleration $\ddot z_\mathrm{c}$ of the contact point.
With again $t' = t + \Delta_t$,
the three equations to be solved are
\begin{eqnarray}
z_\mathrm{c}' & = &  z_\mathrm{c}
+ \Delta_t \dot z_\mathrm{c}
+ \frac{1}{2} \Delta_t^2 \ddot  z_\mathrm{c},
\nonumber \\
\dot  z_\mathrm{c}' & = & \dot  z_\mathrm{c}
+ \Delta_t  \ddot z_\mathrm{c},
\nonumber \\
z_\mathrm{c}' & = & F_z(h',\dot h') /k_\mathrm{eff} .
\end{eqnarray}
Recall $F_z \equiv F_\mathrm{vdW} + F_\mathrm{drain}$.
Because $h' = z_\mathrm{p}' + z_\mathrm{c}'$,
the right hand side of the last equation
is a function of $\ddot  z_\mathrm{c}$.
Hence the correct acceleration is the zero of
\begin{equation}
G(\ddot z_\mathrm{c})
\equiv
 z_\mathrm{c}'(\ddot z_\mathrm{c})
-  F_z(h',\dot h') /k_\mathrm{eff} .
\end{equation}
This is converted into a fixed point problem by defining
\begin{eqnarray}
H(\ddot  z_\mathrm{c})
& \equiv &
\ddot z_\mathrm{c}
- \left( \frac{\partial  G }{\partial \ddot z_\mathrm{c}} \right)^{-1}
G(\ddot z_\mathrm{c})
\\ \nonumber & = &
\ddot z_\mathrm{c}
-
\frac{ z_\mathrm{c} + \Delta_t \dot z_\mathrm{c}
+ 0.5 \Delta_t^2 \ddot z_\mathrm{c}
-  F_z(h',\dot h') /k_\mathrm{eff}
}{ \displaystyle
\frac{ \Delta_t^2 }{2}
-  \frac{\partial F_z(h',\dot h')}{k_\mathrm{eff} \partial h'}
\frac{ \Delta_t^2 }{2}
-  \frac{\partial F_z(h',\dot h')}{k_\mathrm{eff} \partial \dot h'} \Delta_t  }.
\end{eqnarray}
From this, the iteration procedure for the acceleration is
\begin{equation}
\ddot z_\mathrm{c}^{(n+1)} = H(\ddot z_\mathrm{c}^{(n)} ).
\end{equation}
This is quadratically convergent and was found to be quite stable,
including in the contact region where the van der Waals force
is steeply repulsive.

\end{document}